\documentclass[pra,twocolumn,a4paper,showpacs,superscriptaddress]{revtex4}

 \usepackage{latexsym}
 \usepackage{amsmath}
 \usepackage{amsfonts}
 \usepackage{graphicx}
 \usepackage{amssymb}
 \usepackage{subfigure}
 \usepackage{color}
 \usepackage{setspace}
 \usepackage{titlesec}
 \usepackage{mathtools}

 \newcommand{\bc}{\begin{center}}
 \newcommand{\ec}{\end{center}}

\begin{document}

\title{ PT Symmetry, induced mechanical lasing and tunable force sensing in a coupled-mode optically levitated nanoparticle}

\author{Sandeep \surname{Sharma}}
\email{quanta.sandeep@gmail.com}
\affiliation{Department of Physics, Korea Advanced Institute of Science and Technology (KAIST), 
Daejeon 34141, South Korea}
\affiliation{School of Physics and Astronomy, Rochester Institute of Technology, 84 Lomb Memorial Drive,
Rochester, New York, 14623, USA}

\author{A. \surname{Kani}}
%\email{kxmsps@rit.edu}
\affiliation{School of Physics and Astronomy, Rochester Institute of Technology, 84 Lomb Memorial Drive,
Rochester, New York, 14623, USA}

\author{M. \surname{Bhattacharya}}
%\email{mxbsps@rit.edu}
\affiliation{School of Physics and Astronomy, Rochester Institute of Technology, 84 Lomb Memorial Drive,
Rochester, New York, 14623, USA}
\date{\today}

%\pacs{42.50.Gy,42.65.An,42.65.Jx}
%%%%%%%%%%%%%%%%%%%%%%%%%%%%%%%%%%%%%%%%%%
%                                        ABSTRACT                                                                         %
%%%%%%%%%%%%%%%%%%%%%%%%%%%%%%%%%%%%%%%%%%
\begin{abstract}
We theoretically investigate PT symmetry, induced mechanical lasing and force sensing in an optically levitated nanoparticle with coupled oscillation modes. The coupling in the levitated system is created by the modulation of an asymmetric optical potential in the plane transverse to the beam trapping the nanoparticle. We show that such a coupling can lead to PT-symmetric mechanical behavior for experimentally realistic parameters. Further, by examining the phonon dynamics and the second-order coherence of the nanoparticle modes, we determine that induced mechanical lasing is also possible. Finally, we demonstrate that tunable ultra-sensitive force sensing ($\sim zN/\sqrt{Hz}$) can be engineered in the system. Our studies represent an advance in the fields of coherent manipulation of coupled degrees of freedom of levitated mechanical oscillators and their application for sensing.
\end{abstract}
%%%%%%%%%%%%%%%%%%%%%%%%%%%%%%%%%%%%%%%%%%
%%%%%%%%%%%%%%%%%%%%%%%%%%%%%%%%%%%%%%%%%%
\maketitle
%%%%%%%%%%%%%%%%%%%%%%%%%%%%%%%%%%%%%%%%%%
%%%%%%%%%%%%%%%%%%%%%%%%%%%%%%%%%%%%%%%%%%
%        INTRODUCTION                                                        %
%%%%%%%%%%%%%%%%%%%%%%%%%%%%%%%%%%%%%%%%%%
%%%%%%%%%%%%%%%%%%%%%%%%%%%%%%%%%%%%%%%%%%
\section{Introduction}
Coupled optomechanical systems have garnered a lot of attention in the recent years owing to their potential applications in the field of sensing \cite{Hodaei, Djorwe,Zhang,Mao}, quantum information processing \cite{Giovannetti,Mari,Zhang2}, and entanglement \cite{Chen,Wang2}.
 In connection with these applications, a variety of phenomena such as PT (Parity-Time) symmetry \cite{Li,Kepesidis}, quantum synchronization\cite{Liao, Li2}, photon blockade \cite{Li3}, quantum state transfer \cite{Singh, Li4}, etc have been studied in the coupled systems. Most of these proposals involved coupled cavities or coupled clamped nanomechanical resonators, which may be less efficient in obtaining high fidelity entanglement, efficient quantum state transfer, ultra-sensitive sensing, etc. owing to their high decoherence rates \cite{Aspelmeyer}. 
 
 In contrast, realizing a coupled optomechanical configuration using an optically levitated system can be advantageous in overcoming this difficulty as these well-isolated systems are known to have very low decoherence rates \cite{Gieseler,Gieseler2}. Indeed, because they present such advantages, optically levitated systems have been used in applications such as force sensing \cite{Hebestreit,Ranjit}, magnetic sensing \cite{Kumar}, and rotational sensing \cite{Ahn}. Apart from these uses, optically levitated systems have also been used to study squeezing \cite{Ge}, bistability \cite{Ricci}, superposition states \cite{Isart, Sharma}, phonon lasing \cite{Pettit}, etc. However, all these studies are mainly focused on using a single mechanical mode of the optically levitated system. 

Recently, coherent dynamics of a two mode coupled levitated system have been studied by Frimmer et.al. \cite{Frimmer2}. In their work, they have coupled the two transverse modes of an optically levitated nanoparticle by modulating an asymmetric trap potential in the transverse plane of the system. By exploiting this coupling, they have implemented cooling of the transverse modes in this system. However, to the best of our knowledge, apart from the above mentioned studies on coherent dynamics, not much has been explored about this coupled nanoparticle system. It would be interesting to see, for example, how effective the coupled levitated systems could be in demonstrating two-mode phenomena such as PT symmetry, entanglement, synchronization, and quantum state transfer.

In this work we propose to use the coupled-mode levitated nanoparticle as a tool box for studying the aforementioned phenomena, see Fig.~\ref{fig:fig1}. The coupling is introduced by manipulating the trap potential in the plane transverse to the trapping beam, which propagates along the $z$ axis, as proposed by Frimmer et. al. \cite{Frimmer2}. The modulation of the asymmetric potential results in the coupling of the two transverse modes {\it x} and {\it y}. By suitably amplifying and cooling the mechanical motion in the two modes, respectively, we find that the system can be driven into the PT-symmetric regime. Numerical studies of phonon dynamics and second-order coherence imply that the system remains in a thermal state in this regime. However, we show that when one of the modes is a lasing mode, then the coupling drives the other modes also towards lasing. Finally, we exploit the mode-mode coupling to achieve highly tunable ultrasensitive force sensing. We find that in the strong coupling regime, weak forces can be measured with high sensitivity at different frequencies tunably using this system. 

In our simulations we use realistic experimental parameters and take into account relevant sources of noise and dissipation. Our work on analyzing various implications of modal coupling in an optically levitated nanoparticle opens up new possibilities for coherent manipulation and sensing using these highly isolated systems. In the remainder of this paper, we introduce our theoretical formulation in Section \ref{TF}, present
the results and discussion in Section \ref{RD}, and our conclusions in Section \ref{Con}.
%the efficiency of the above mentioned applications 
%%%%%%%%%%%%%%%%%%%%%%%%%%%%%%%%%%%%%%%%%%
%%%%%%%%%%%%%%%%%%%%%%%%%%%%%%%%%%%%%%%%%%
%                                        Model system and its basic equations                                 %
%%%%%%%%%%%%%%%%%%%%%%%%%%%%%%%%%%%%%%%%%%
\section{Theoretical Formulation}
\label{TF}
\subsection{Model}
\begin{figure}[t]
\includegraphics[height=4.5cm, width=8.5cm]{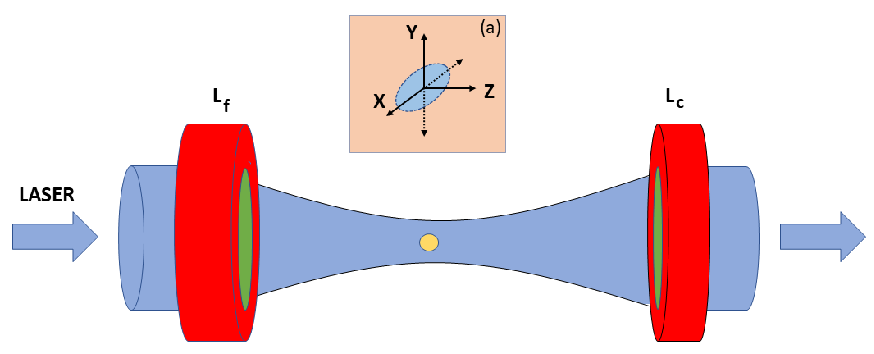}%
 \caption[]{A schematic diagram for a levitated nanoparticle system considered in this work. A focusing lens $L_{f}$ is used to trap the nanoparticle at the focus of the trap laser and the information about particle position is collected by the lens $L_{c}$ to be processed. The figure labeled as (a) in the inset shows an asymmetric potential created in the transverse $x-y$ plane at the focus of the trap laser.}
\label{fig:fig1}%
\end{figure}

We consider a single dielectric nanoparticle of mass {\it m} optically trapped in the potential created by a focused Gaussian beam under high vacuum, as shown in Fig.~\ref{fig:fig1}. The created optical potential is harmonic to a good approximation around the focus of the Gaussian beam. Hence the trapped nanoparticle can be considered as a harmonic oscillator, whose three modes of oscillation are decoupled for small amplitudes, along the three different axes {\it x}, {\it y} and {\it z}, respectively \cite{Frimmer1}. 

In this paper, we concentrate on the particle dynamics in the {\it x-y} plane only, while freezing the particle motion along {\it z} direction using feedback cooling. Further, the coupling of the motion along {\it x} and {\it y} direction is achieved by modulating the polarization of the trap laser beam using an electro-optic modulator \cite{Frimmer2}. This modulation varies the asymmetric potential in the {\it x-y} plane, thereby coupling the modes. In order to study the dynamics of the coupled-mode system, we write the master equation for this system \cite{Rodenburg,Pettit} as 
%%%%%%%%%%%%%%%%%%%%%%%%%%%%%%%%%%%%%%%%%%%%%
\begin{align}
\label{Eq_1}
\dot{\rho}_{m}&=-i\sum_{j=x,y}^{} \left(\omega_{j}[a^{\dagger}_{j}a_{j},\rho_{m}]+\left(\frac{\gamma_{gj}-\gamma_{aj}}{2}\right)[Q_{j},\{P_{j},\rho_{m}\}]\right) \nonumber \\
                           &-\sum_{j=x,y}^{} \left(\frac{D_{tj}}{2}\mathcal{D}[Q_{j}]\rho_{m}-\frac{D_{j}}{2}\mathcal{D}[P_{j}]\rho_{m}\right) \nonumber \\
                           &-\sum_{j=x,y}^{} \bigg(i\gamma_{cj}[Q^{3}_{j},\{P_{j},\rho_{m}\}] -\Gamma_{cj}\mathcal{D}[Q^{3}_{j}]\rho_{m}\bigg) \nonumber \\
	                &-i\frac{\kappa \delta}{m\sqrt{\omega_{x}\omega_{y}}}\cos(\omega_{r}t)[Q_{x}Q_{y},\rho_{m}],
\end{align}
%%%%%%%%%%%%%%%%%%%%%%%%%%%%%%%%%%%%%%%%%%%%% 
where, $\rho_{m}$ is the density matrix for the two-dimensional coupled system and the dimensionless position and momentum operators for the nanoparticle are denoted as $Q_{j}$ and $P_{j}$, respectively. Here, $j \in \{x,y\}$ and symbolizes the two transverse modes {\it x} and {\it y}, respectively. Further, the commutators and anti-commutators are represented by square $[~]$ and curly brackets $\{~\}$, respectively. The mechanical modes are also represented by phonon creation ($a^{\dagger}_{j}$) and annihilation $(a_{j})$ operators that obey the bosonic commutation relation $[a,a^{\dagger}]=1$. The Lindblad superoperator $\mathcal{D}[\mathcal{O}]$ is defined as 
%%%%%%%%%%%%%%%%%%%%%%%%%%%%%%%%%%%%%%%%%%%%%\textcolor{blue}{}
\begin{align}
\label{Eq_1a}
\mathcal{D}[\mathcal{O}]=\mathcal{O}^{\dagger}\mathcal{O}\rho+\rho\mathcal{O}^{\dagger}\mathcal{O}-2\mathcal{O}^{\dagger}\rho\mathcal{O}.
\end{align}
%%%%%%%%%%%%%%%%%%%%%%%%%%%%%%%%%%%%%%%%%%%%% 
The first term on the right hand side of Eq.~(\ref{Eq_1}) corresponds to the harmonic motion of the system with frequencies $\omega_{x}$ and $\omega_{y}$ in the directions {\it x} and {\it y}, respectively. The second term represents damping (anti-damping) due to the surrounding gas (linear feedback amplification) of the system with a rate $\gamma_{gj}$($\gamma_{aj}$) \cite{Pettit}. However, in our case, linear amplification of only one of the modes is of paramount interest, namely the {\it y} mode of the coupled system. The third and fourth terms depict the momentum diffusion (due to photon scattering) and position diffusion (due to photon scattering and amplification backaction) with rates $D_{tj}$ and $D_{j}$, respectively. The fifth and sixth term appear due to nonlinear feedback and its concomitant back-action effect, respectively. The rates of feedback cooling and cooling back-action are given by $\gamma_{cj}$ and $\Gamma_{cj}$, respectively. The last term in Eq.~(\ref{Eq_1}) arises due to the coupling of the two transverse modes {\it x} and {\it y} modes. The coupling is introduced by periodically rotating the asymmetric potential around {\it z-}axis by a small angle $\delta$ at the frequency $\omega_{r}(\approx~ \omega_{y}-\omega_{x})$. Further, $\kappa \simeq m(\omega^{2}_{y}-\omega^{2}_{x})/2~$ denotes the change in trap stiffness due to the asymmetry in the trapping potential. Throughout this paper, we will analyze the effect of this coupling on the dynamics of the system, and its relation to the realization of various interesting and useful phenomena.
%#################################################################%
%%%%%%%%%%%%%%%%%%%%%%%%%%%%%%%%%%%%%%%%%%%%%%%%%%%%%%%%%%%%%%%%%%% $\gamma_{g}=0.1\omega_{0}$,  $\gamma_{g}=5\omega_{0}$,  $\gamma_{g}=0.1\omega_{0}$, and kerr parameter $U = 0.2\omega_{0}$,  $\gamma_{g}=5\omega_{0}$, and kerr parameter $U = 0.2\omega_{0}
%         FIGURE-1                                                                                                                                                                                      
%%%%%%%%%%%%%%%%%%%%%%%%%%%%%%%%%%%%%%%%%%%%%%%%%%%%%%%%%%%%%%%%%%%%
%%%%%%%%%%%%%%%%%%%%%%%%%%%%%%%%%%%%%%%%%%%%%%%%%%%%%%%%%%%%%%%%%%%%%%
%%%%%%%%%%%%%%%%%%%%%%%%%%%%%%%%%%%%%%%%%%%%%%%%%%%
\subsection{Quantum Langevin equations}
%#################################################################%
In order to study the effect of coupling on the position dynamics of both the modes of the coupled system, we use the full quantum Langevin equations of motion including all quantum fluctuations for the system, derived from Eq.~(\ref{Eq_1}). The complete equations of motion for the coupled system are given as
\begin{align}
\label{Eq_2a}
\dot{Q}_{x}&=\omega_{x}P_{x}, \\
\dot{P}_{x}&=-\omega_{x}Q_{x}-2(\gamma_{gx}+24\gamma_{cx}Q^{2}_{x})P_{x}\nonumber \\
                     &-\frac{\kappa \delta}{m\sqrt{\omega_{x}\omega_{y}}}\cos(\omega_{r}t) Q_{y}, \nonumber \\
                     &+\sqrt{\frac{2K_{B}T\gamma_{gx}}{\hbar\omega_{x}}}\xi_{T}+\sqrt{D_{tx}}\xi_{Fa}+12Q^{2}_{x}\sqrt{\frac{\Gamma^{2}_{cx}}{\gamma_{cx}}}\xi_{Fc}, \\
\dot{Q}_{y}&=\omega_{y}P_{y}, \\
\dot{P}_{y}&=-\omega_{y}Q_{y}-2(\gamma_{gy}-\gamma_{ay}+24\gamma_{cy}Q^{2}_{y})P_{y}\nonumber \\
                     &-\frac{\kappa \delta}{m\sqrt{\omega_{x}\omega_{y}}}\cos(\omega_{r}t)  Q_{x}, \nonumber \\
                     &+\sqrt{\frac{2K_{B}T\gamma_{gy}}{\hbar\omega_{y}}}\xi_{T}+\sqrt{D_{ty}}\xi_{Fa}+12Q^{2}_{y}\sqrt{\frac{\Gamma^{2}_{cy}}{\gamma_{cy}}}\xi_{Fc},
\label{Eq_2b}
\end{align}
%%%%%%%%%%%%%%%%%%%%%%%%%%%%%%%%%%%%%%%%%%%%%%%%%%%%%%%%%%%%%%%%%%%%%
where the correlations corresponding to the zero-mean noise from environment, feedback amplification and cooling are represented as  $\langle \xi_{T}(t)\xi_{T}(t') \rangle = \delta(t-t')$, $\langle \xi_{Fa}(t)\xi_{Fa}(t') \rangle = \delta(t-t')$, and $\langle \xi_{Fc}(t)\xi_{Fc}(t') \rangle = \delta(t-t')$, respectively.

Now, in order to further simplify the above equations of motion, we use the following transformations
%%%%%%%%%%%%%%%%%%%%%%%%%%%%%%%%%%%%%%%%%%%%%%%%%%%%%%%%%%%%%%%%%%%%% 
\begin{align}
\label{Eq_2ab}
Q_{x}&=Q_{0} {\rm Re} \left\{a_{x}(t)~\exp\left[i\left(\omega_{0}-\frac{\omega_{r}}{2}\right)t\right]
	\right\}, \\
Q_{y}&=Q_{0} {\rm Re}\left\{a_{y}(t)~\exp\left[i\left(\omega_{0}+\frac{\omega_{r}}{2}\right)t\right]\right\},
\label{Eq_2ac}
\end{align}
%%%%%%%%%%%%%%%%%%%%%%%%%%%%%%%%%%%%%%%%%%%%%%%%%%%%%%%%%%%%%%%%%%%%%%
where $Q_{0}$ is the initial amplitude of the oscillation modes, $\omega_{0}=\sqrt{\omega_{x}^{2}+\omega_{y}^{2}}/2$ is the carrier frequency  and the complex amplitudes for the oscillation modes along $x$-mode and $y$-mode are represented as $a_{x}$ and $a_{y}$, respectively.

Further, in our present work we are interested in studying the dynamics of mean position for both the modes in the classical limit \cite{Agarwal}. Hence, we neglect the quantum fluctuations and use the slowly varying amplitude approximation and the rotating wave approximation to obtain the mean-value equations for the oscillation amplitudes of the coupled system as 
%%%%%%%%%%%%%%%%%%%%%%%%%%%%%%%%%%%%%%%%%%%%%%%%%%%%%%%%%%%%%%%%%%%%%%%
\begin{align}
\label{Eq_3a}
\langle\dot{a}_{x}\rangle&=-\frac{i}{2}(\Delta-i\Gamma_{x})\langle a_{x}\rangle-i\beta_{x}\langle a_{y}\rangle,  \\
\langle\dot{a}_{y}\rangle&=-i\beta_{y}\langle a_{x}\rangle+\frac{i}{2}(\Delta+i\Gamma_{y})\langle a_{y}\rangle.
\label{Eq_3b}
\end{align}
%%%%%%%%%%%%%%%%%%%%%%%%%%%%%%%%%%%%%%%%%%%%%%%%%%%%%%%%%%%%%%%%%%%%%%%%
where $\Delta=\omega_{1}-\omega_{r}$, $\omega_{1}=(\omega_{y}^{2}-\omega_{x}^{2})/2\omega_{0}$, $\Gamma_{x}=2(\gamma_{gx}+24\gamma_{cx}\langle a^{2}_{x}\rangle)$, $\Gamma_{y}=2(\gamma_{gy}-\gamma_{ay}+24\gamma_{cy}\langle a^{2}_{y}\rangle)$, $\beta_{x}=\omega_{3}\sqrt{\omega_{x}/\omega_{y}}$, $\beta_{y}=\omega_{3}\sqrt{\omega_{y}/\omega_{x}}$, and $\omega_{3}=[\delta(\omega_{y}^{2}-\omega_{x}^{2})]/2\omega_{0}$.
Next, we rewrite the above dynamical equations into the following matrix form:
%%%%%%%%%%%%%%%%%%%%%%%%%%%%%%%%%%%%%%%%%%%%%%%%%%%%%%%%%%%%%%%%%%%%%%%%%%%%
\begin{gather}
i
\begin{bmatrix}
\langle\dot{a}_{x}\rangle\\
\\
\langle\dot{a}_{y}\rangle\\
\end{bmatrix}
= H
\begin{bmatrix}
\langle{a}_{x}\rangle\\
\\
\langle{a}_{y}\rangle\\
\end{bmatrix}
\label{Eq_3bc}
\end{gather}
%%%%%%%%%%%%%%%%%%%%%%%%%%%%%%%%%%%%%%%%%%%%%%%%%%%%%%%%%%%%%%%%%%%%%%%%%%%%
where $H$ is the non-Hermitian Hamiltonian for the coupled mode system and is described as
%%%%%%%%%%%%%%%%%%%%%%%%%%%%%%%%%%%%%%%%%%%%%%%%%%%%%%%%%%%%%%%%%%%%%%%%%%%%
\[
H =
\begin{bmatrix}
\frac{1}{2}(\Delta-i\Gamma_{x}) & \beta_{x}\\
\\
\beta_{y} & -\frac{1}{2}(\Delta+i\Gamma_{y})\\
\end{bmatrix}\]
%%%%%%%%%%%%%%%%%%%%%%%%%%%%%%%%%%%%%%%%%%%%%%%%%%%%%%%%%%%%%%%%%%%%%%%%%%%%
In the later section, we study the effect of coupling on the eigenvalues of this Hamiltonian as it is a key step to establish PT symmetry in the coupled levitated system.
%%%%%%%%%%%%%%%%%%%%%%%%%%%%
\subsection{Phonon dynamics}        %%%%%%%%%%%
%%%%%%%%%%%%%%%%%%%%%%%%%%%%
In this subsection, we focus on the dynamics of the phonon population for the coupled-mode system. The equation representing phonon dynamics is constructed using  $\langle\dot{N}\rangle=$~Tr$[N\rho]$ along with Eq.~(\ref{Eq_3a}) and ~(\ref{Eq_3b}). We then make use of the factorization assumptions $ \langle a_{i} a_{j}\rangle \sim  \langle a_{i}\rangle \langle a_{j} \rangle $ \cite{Agarwal} which hold in the classical (mean-field) limit, and write the dynamical equations for the evolution of mean phonon number of both modes as
%#################################################################%
\begin{align}
\label{Eq_4a}
\langle\dot{N}_{x}\rangle&=-\Gamma_{x} \langle N_{x}\rangle+(D_{tx}-6\gamma_{cx})\nonumber\\
&+ i \beta_{x}[\langle a_{y}^{\dagger}\rangle \langle a_{x}\rangle-\langle a_{x}^{\dagger}\rangle \langle a_{y}\rangle], \\
\langle\dot{N}_{y}\rangle&=-\Gamma_{y} \langle N_{y}\rangle+(\gamma_{ay}+D_{ty}-6\gamma_{cy})\nonumber\\
&+ i \beta_{y}[\langle a_{x}^{\dagger}\rangle \langle a_{y}\rangle-\langle a_{y}^{\dagger}\rangle \langle a_{x}\rangle].
\label{Eq_4b}
\end{align}
%%%%%%%%%%%%%%%%%%%%%%%%%%%%%%%%%%%%%%%%%%%%%%%%%%%%%%%%%%%%%%%%%%%%%
Study of phonon dynamics is very important as a saturation-like behavior of the phonon number can indicate lasing action in the system \cite{Pettit}. However, such behavior in the phonon dynamics is necessary but not sufficient for validating lasing action. Therefore, we also study the second-order coherence for the coupled-mode system in the following section. 
%%%%%%%%%%%%%%%%%%%%%%%%%%%%%%%%%%%%%%%%%%%%%%%%%%%%%%%%%%%%%%%%%%% $\gamma_{g}=0.1\omega_{0}$,  $\gamma_{g}=5\omega_{0}$,  $\gamma_{g}=0.1\omega_{0}$, and kerr parameter $U = 0.2\omega_{0}$,  $\gamma_{g}=5\omega_{0}$, and kerr parameter $U = 0.2\omega_{0}
%         FIGURE-2                                                                                                                                                                                       
%%%%%%%%%%%%%%%%%%%%%%%%%%%%%%%%%%%%%%%%%%%%%%%%%%%%%%%%%%%%%%%%%%%%
\begin{figure}[t!]
 \centering
 \subfigure[][]{%
\label{fig:fig2a}%
\includegraphics[height=3.9cm, width=3.9cm]{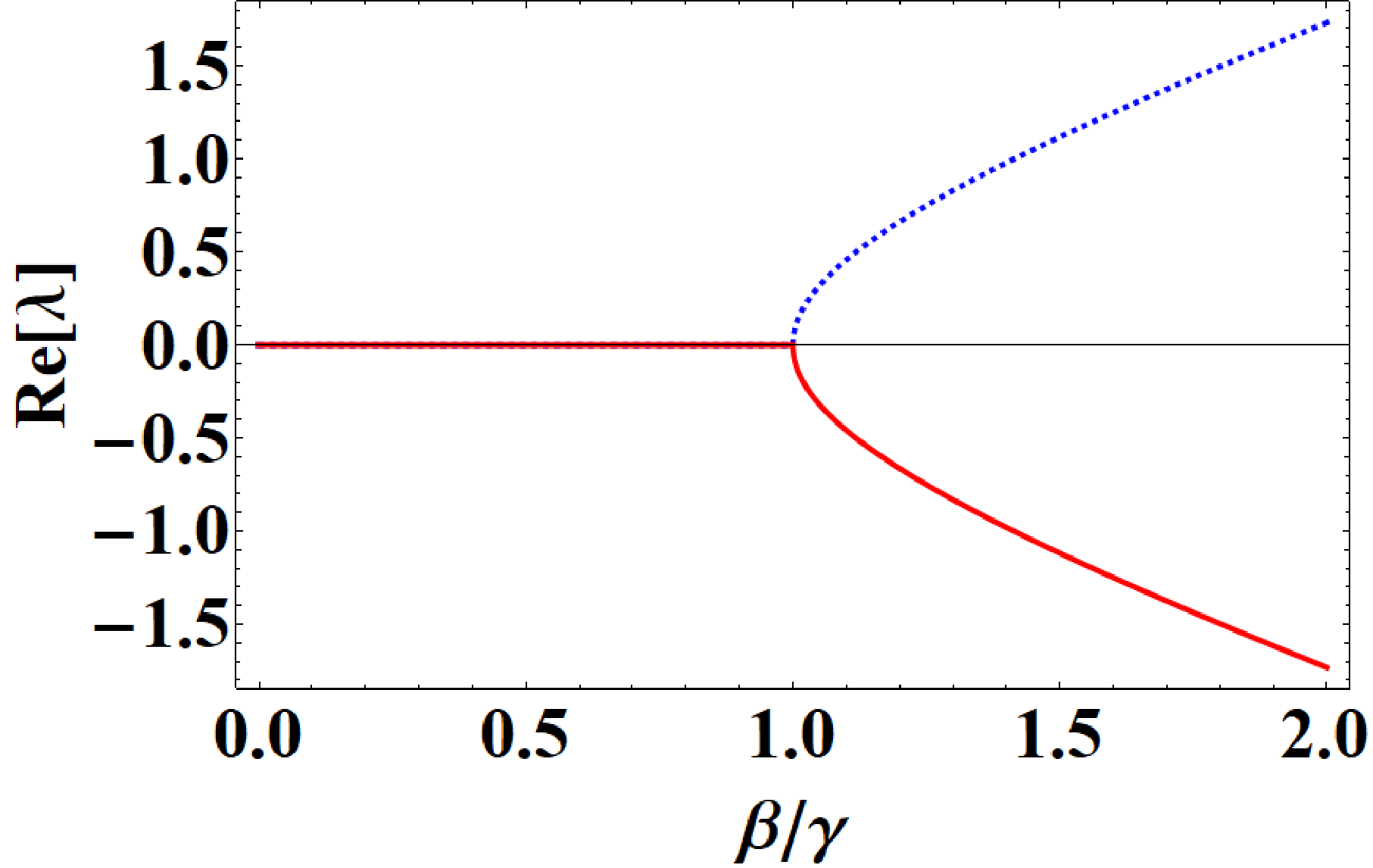}}%
\hspace{8pt}%
\subfigure[][]{%
\label{fig:fig2b}%
\includegraphics[height=3.9cm, width=3.9cm]{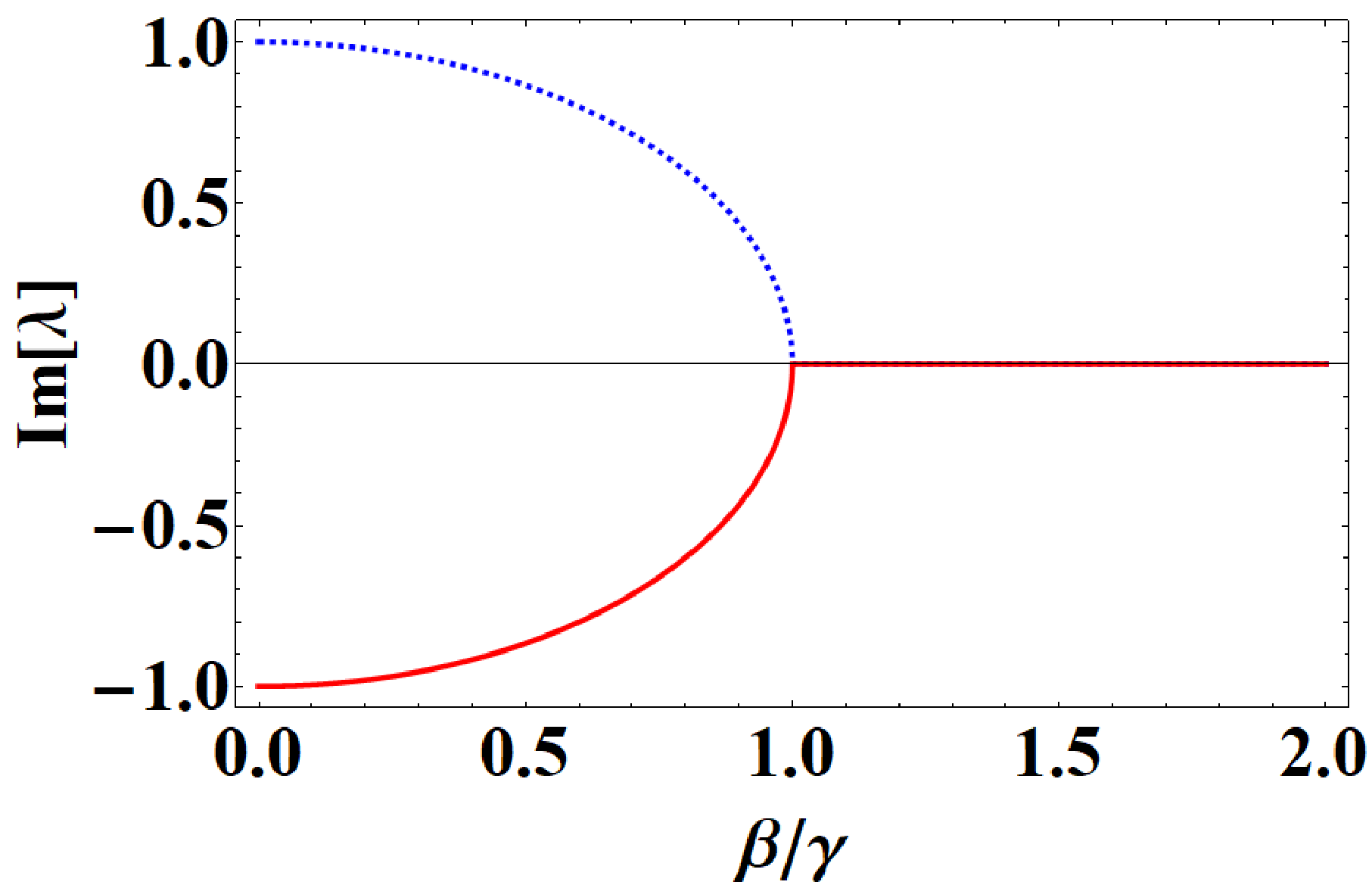}}
 \caption[]{Shows PT symmetry in coupled mode levitated nanoparticle system. Panels (a) and (b) show the variation of real and imaginary part of the eigenvalues $\lambda_{+}$(blue dashed), and $\lambda_{-}$(red solid) with coupling strength $\beta$, respectively. Parameters are $\omega_{x}=130$ KHz, $\omega_{y}=160$ KHz, $\gamma_{gx}=\gamma_{gy}=\gamma=0.06$ Hz, $\gamma_{ay}=0.12$ Hz, $\gamma_{cx}=\gamma_{cy}=0.0$ Hz, and $\Delta=0$ Hz.}
 \label{fig:fig2}
 \end{figure}
%%%%%%%%%%%%%%%%%%%%%%%%%%%%%%%%%%%%%%%%%%%%%%%%%%%%%%%%%%%%%%%%%%%%%%
%%%%%%%%%%%%%%%%%%%%%%%%%%%%%%%%%%%%%%%%%%%%%%%%%%%%%%%%%%%%%%%%%%% $\gamma_{g}=0.1\omega_{0}$,  $\gamma_{g}=5\omega_{0}$,  $\gamma_{g}=0.1\omega_{0}$, and kerr parameter $U = 0.2\omega_{0}$,  $\gamma_{g}=5\omega_{0}$, and kerr parameter $U = 0.2\omega_{0}
%         FIGURE-3                                                                                                                                                                                      
%%%%%%%%%%%%%%%%%%%%%%%%%%%%%%%%%%%%%%%%%%%%%%%%%%%%%%%%%%%%%%%%%%%%
\begin{figure}[t!]
 \centering
 \subfigure[][]{%
\label{fig:fig3a}%
\includegraphics[height=3.9cm, width=3.9cm]{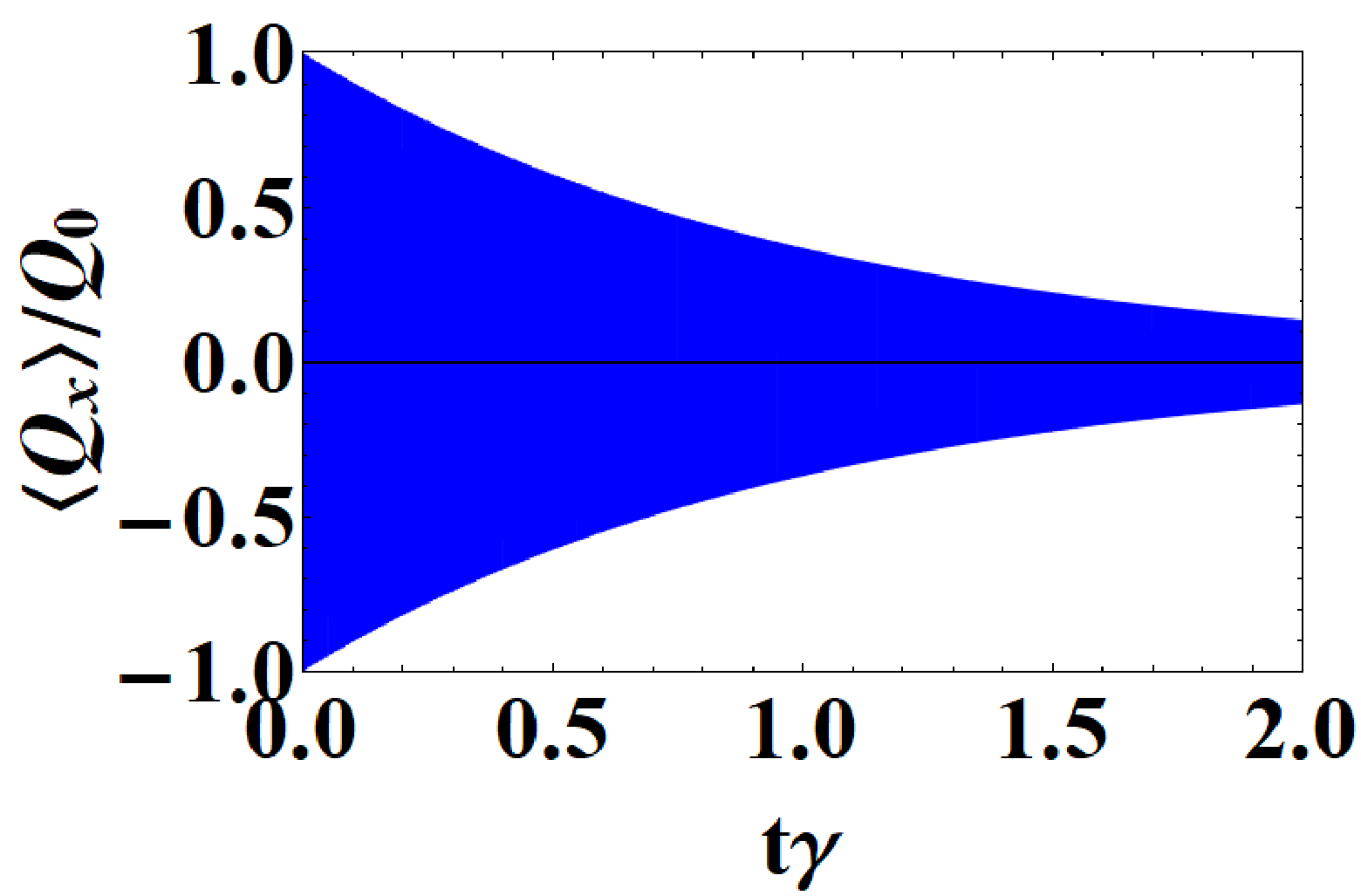}}%
\hspace{8pt}%
\subfigure[][]{%
\label{fig:fig3b}%
\includegraphics[height=3.9cm, width=3.9cm]{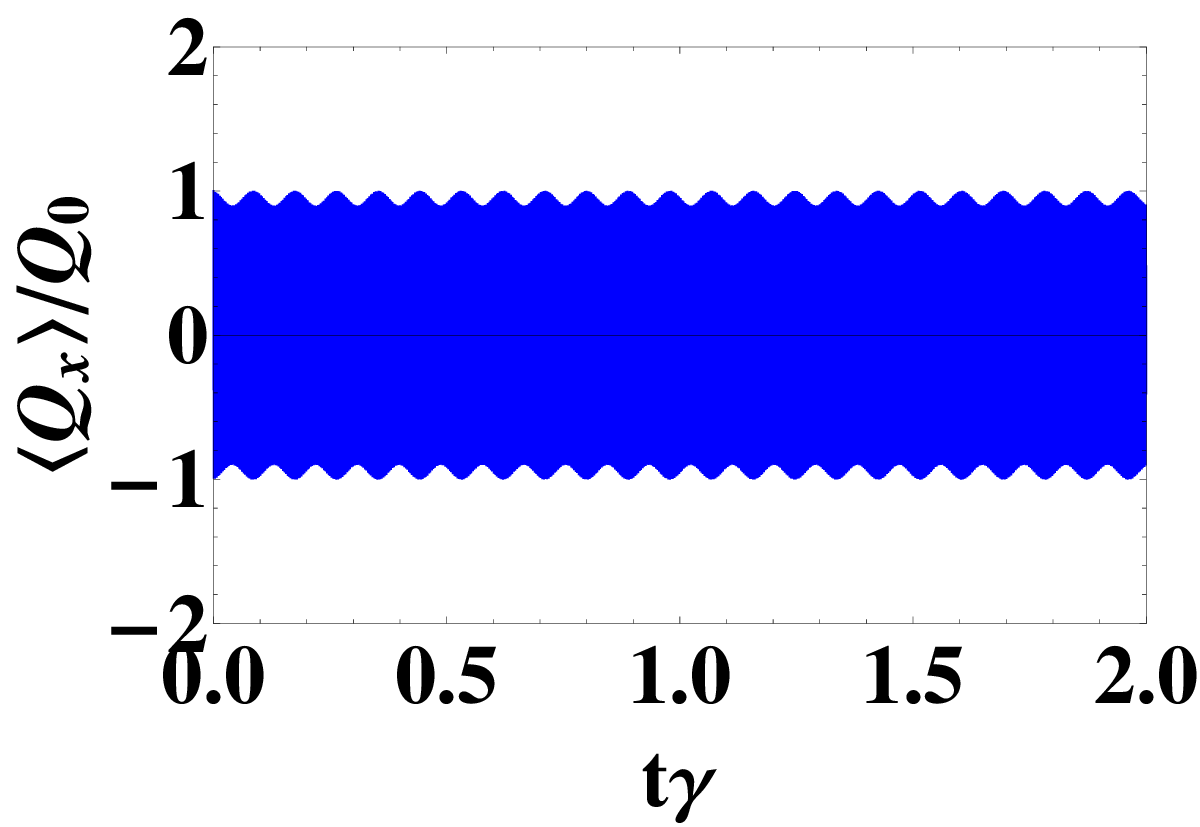}} \\
\hspace{8pt}%
\subfigure[][]{%
\label{fig:fig3c}%
\includegraphics[height=3.9cm, width=3.9cm]{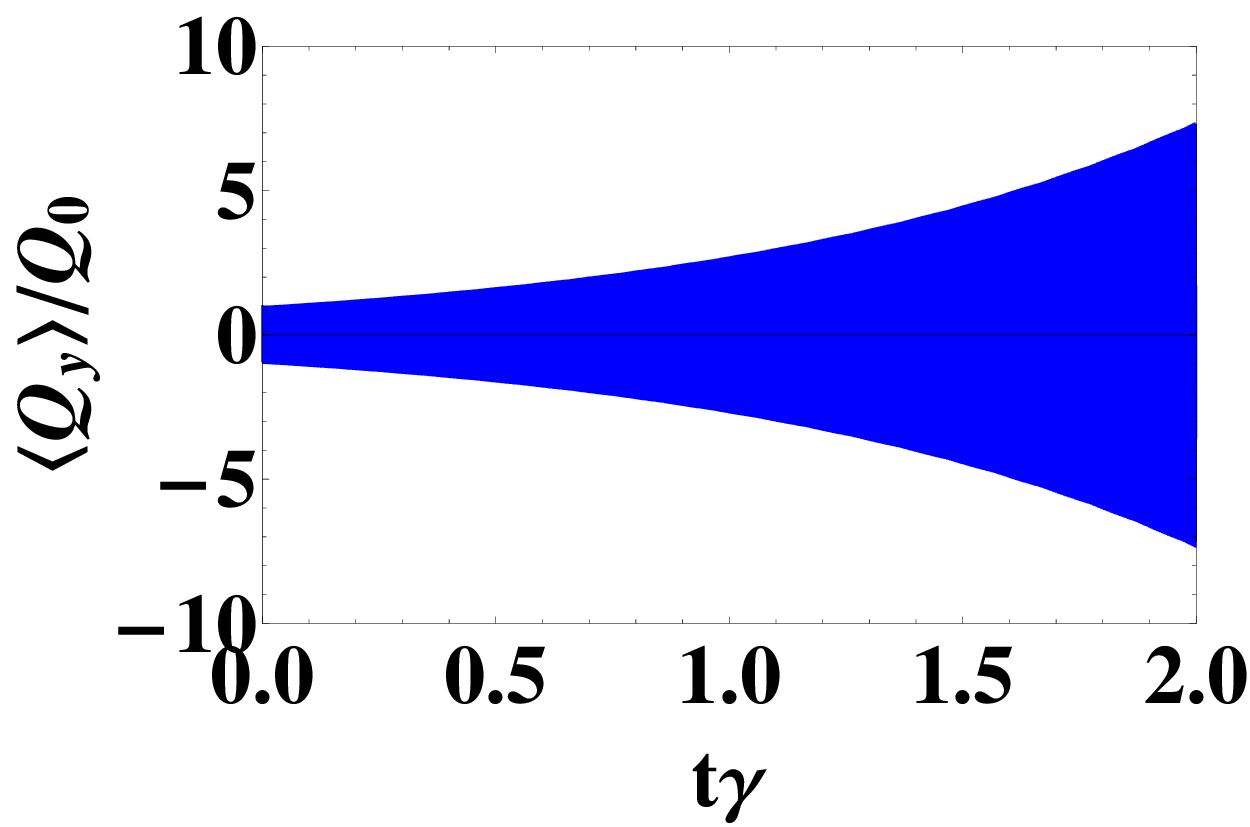}} 
\hspace{8pt}%
\subfigure[][]{%
\label{fig:fig3d}%
\includegraphics[height=3.9cm, width=3.9cm]{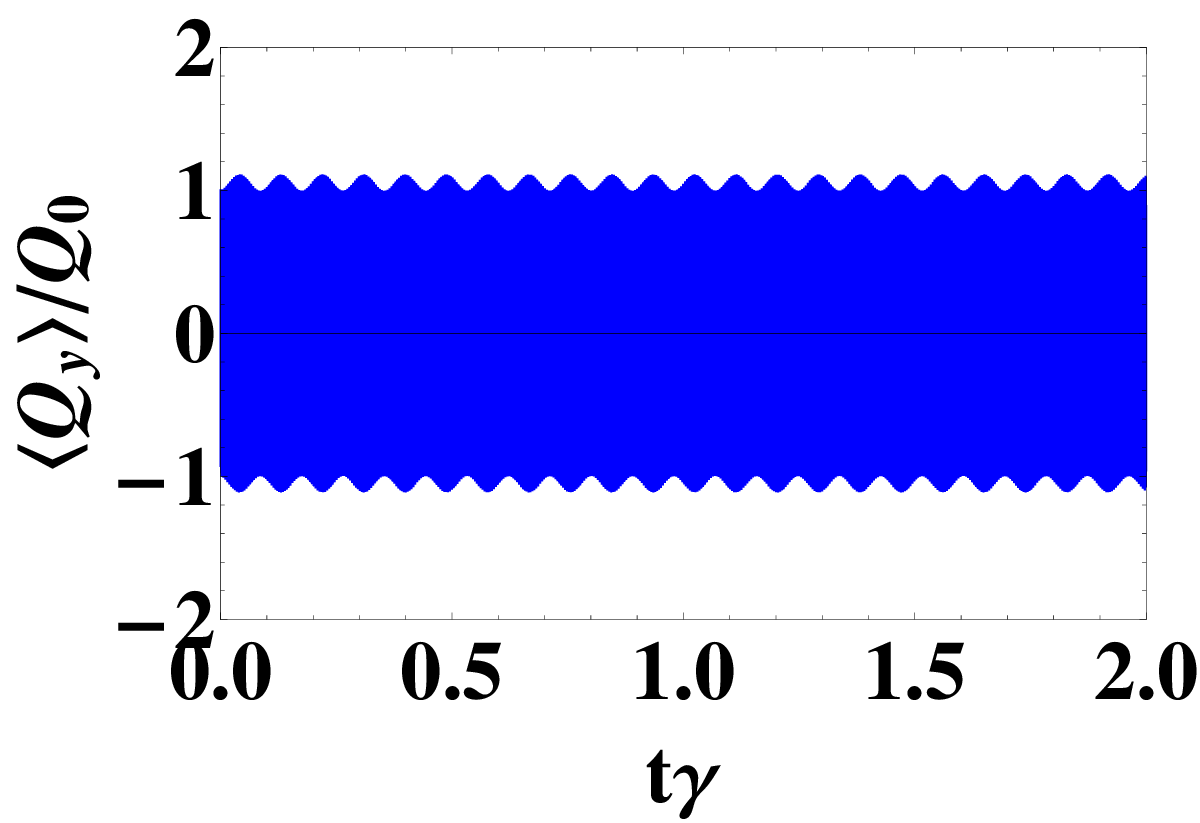}} 
 \caption[]{The evolution of oscillation amplitudes for a coupled-mode levitated nanoparticle. Panels (a) and (b) show the dynamics of the $x$-mode for $\delta=0$ and  $\delta=10^{-4}$, respectively. Panels (c) and (d) show the dynamics of the $y$-mode for $\delta=0$ and  $\delta=10^{-4}$ , respectively. The remaining parameters are $\omega_{x}=130$ KHz, $\omega_{y}=160$ KHz, $\gamma_{gx}=\gamma_{gy}=\gamma=0.06$ Hz, $\gamma_{ay}=0.12$ Hz, $\gamma_{cx}=\gamma_{cy}=0.0$ Hz, $\Delta=10^{-2}$ Hz, and $Q_{0}=3\times 10^{2}$.}
\label{fig:fig3}
\end{figure}
%%%%%%%%%%%%%%%%%%%%%%%%%%%%%%%%%%%%%%%%%%%%%%%%%%%%%%%%%%%%%%%%%%%%%%
%%%%%%%%%%%%%%%%%%%%%%%%%%%%%%%%%%%%%%%%%%%%%%%%%%%%%%%%%%%%%%%%%%% $\gamma_{g}=0.1\omega_{0}$,  $\gamma_{g}=5\omega_{0}$,  $\gamma_{g}=0.1\omega_{0}$, and kerr parameter $U = 0.2\omega_{0}$,  $\gamma_{g}=5\omega_{0}$, and kerr parameter $U = 0.2\omega_{0}
%         FIGURE-4                                                                                                                                                                                      
%%%%%%%%%%%%%%%%%%%%%%%%%%%%%%%%%%%%%%%%%%%%%%%%%%%%%%%%%%%%%%%%%%%%
\begin{figure}[t!]
 \centering
 \subfigure[][]{%
\label{fig:fig4a}%
\includegraphics[height=3.9cm, width=3.9cm]{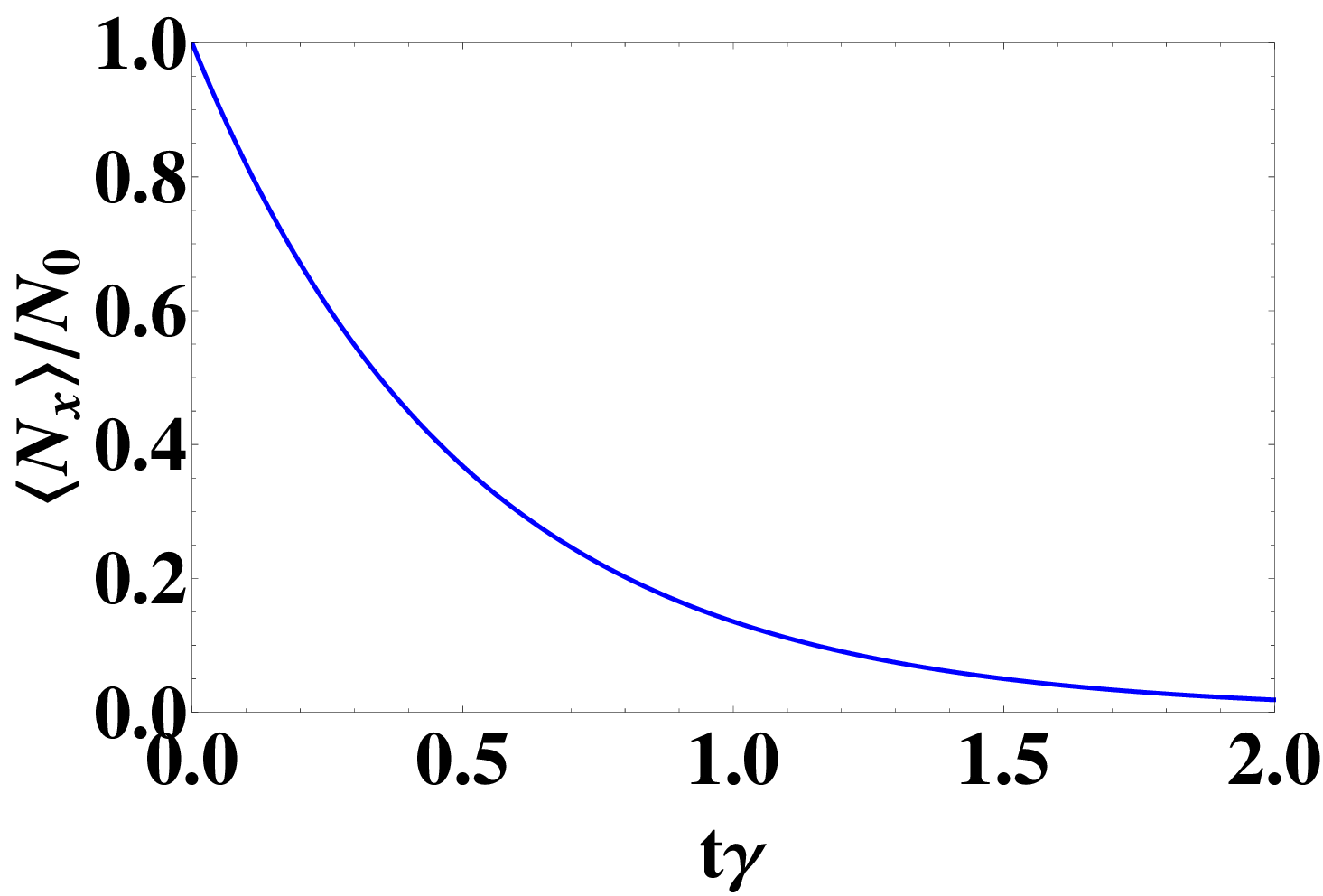}}%
\hspace{8pt}%
\subfigure[][]{%
\label{fig:fig4b}%
\includegraphics[height=3.9cm, width=3.9cm]{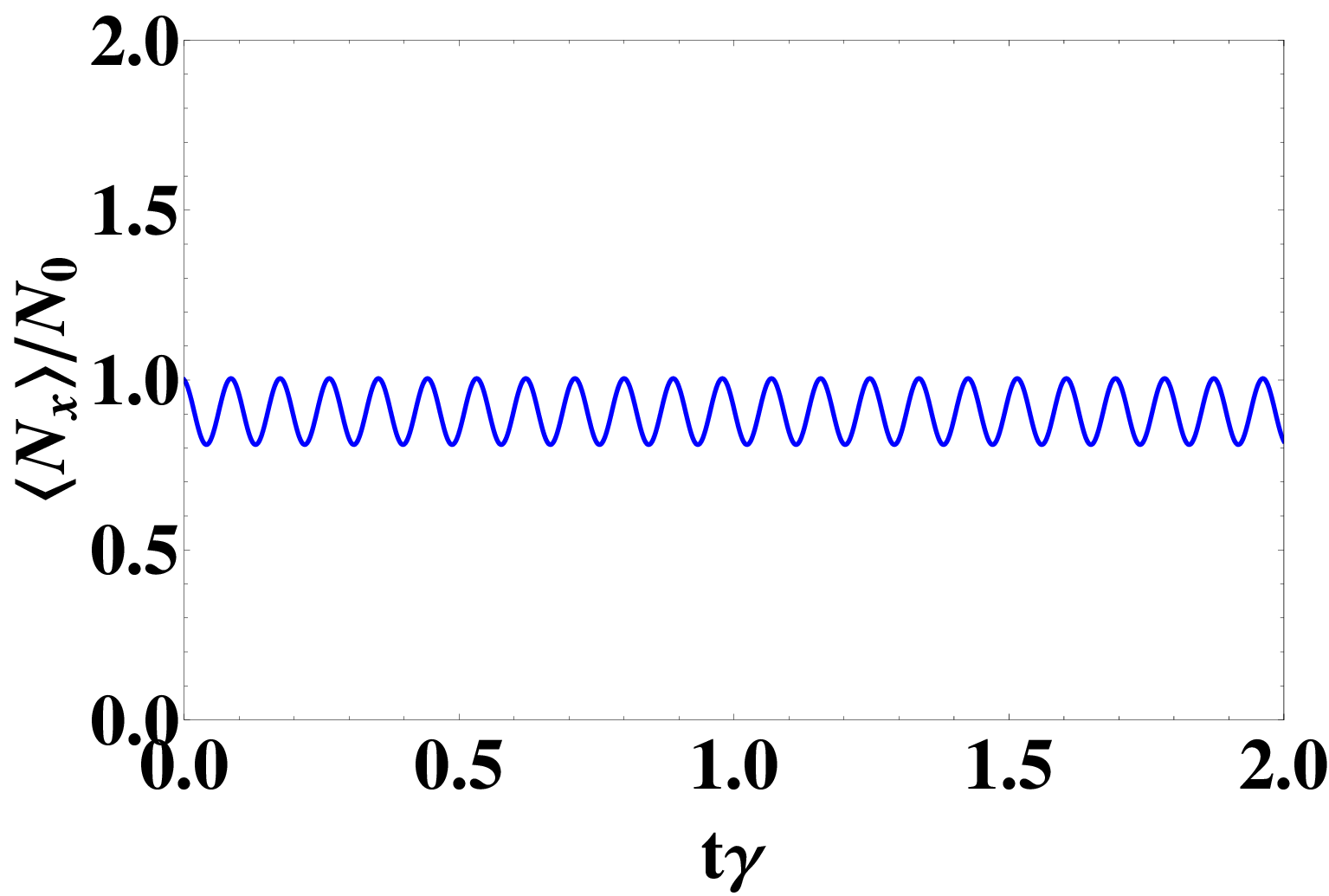}} \\
\hspace{8pt}%
\subfigure[][]{%
\label{fig:fig4c}%
\includegraphics[height=3.9cm, width=3.9cm]{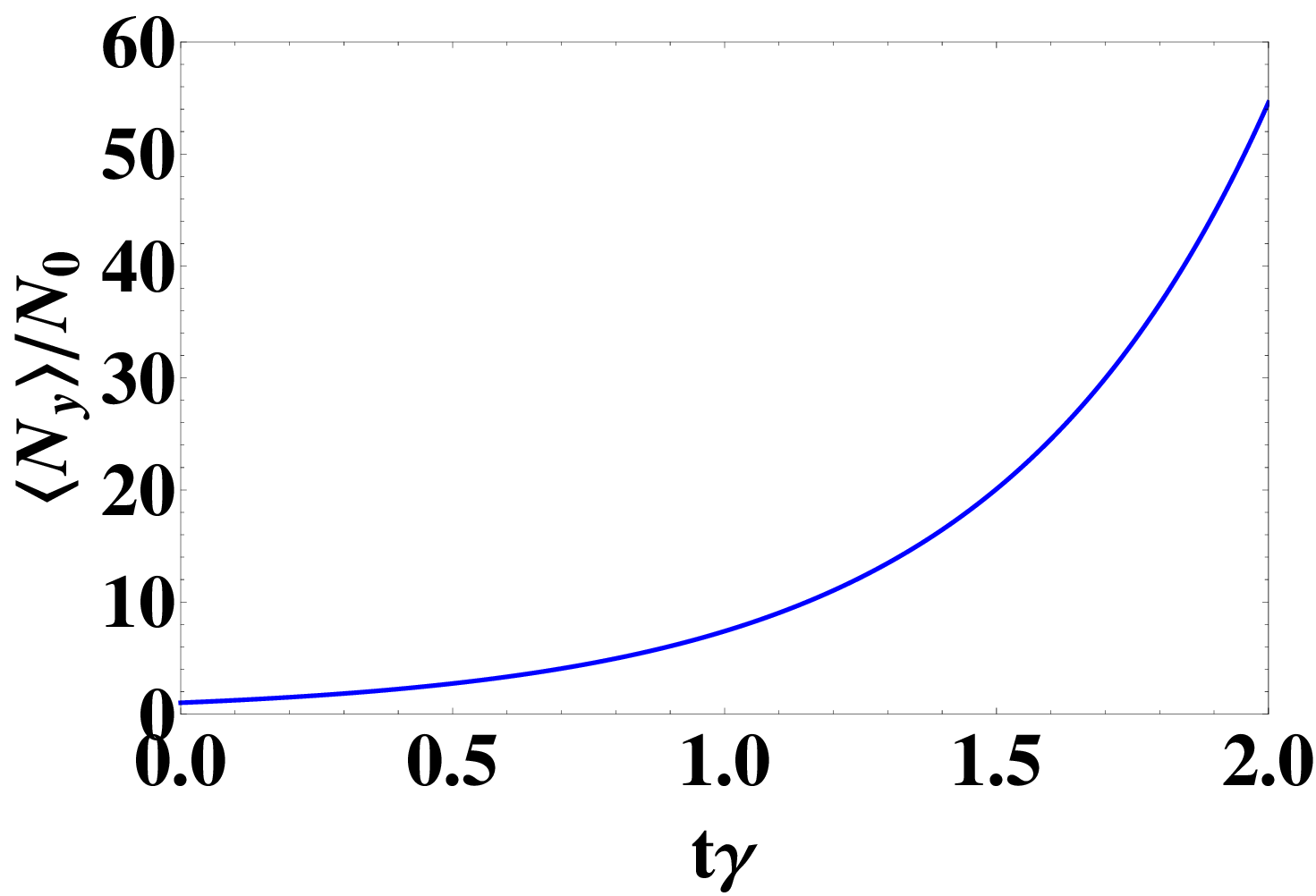}} 
\hspace{8pt}%
\subfigure[][]{%
\label{fig:fig4d}%
\includegraphics[height=3.9cm, width=3.9cm]{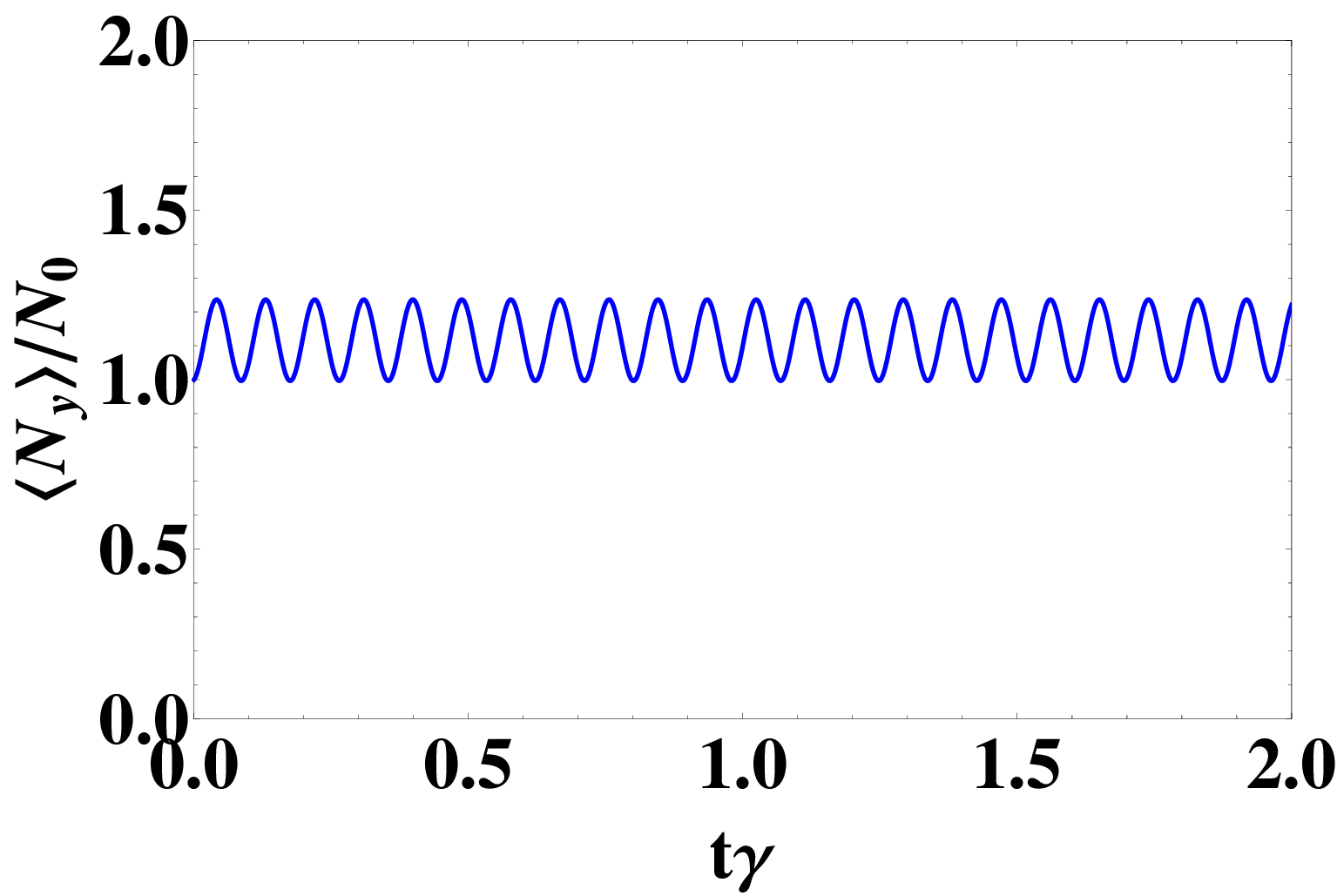}} 
 \caption[]{The time evolution of the phonon population for a coupled-mode levitated nanoparticle. Panels (a) and (b) show the phonon dynamics of the $x$-mode for $\delta=0$ and  $\delta=10^{-4}$, respectively. Panels (c) and (d) show the dynamics of the $y$ mode for $\delta=0$ and  $\delta=10^{-4}$, respectively. Here $N_{0}=10^{5}$, while other parameters are the same as in Fig~\ref{fig:fig3}.}
\label{fig:fig4}
\end{figure}
%%%%%%%%%%%%%%%%%%%%%%%%%%%%%%%%%%%%%%%%%%%%%%%%%%%%%%%%%%%%%%%%%%%%%% parameter $U = 0.2\omega_{0}
%%%%%%%%%%%%%%%%%%%%%%%%%%%%%%%%%%%%%%%%%%%%%%%%%
%%%%%%%%%%%%%%%%%%%%%%%%%%%%%%%%%%%%%%%%
\subsection{Second order coherence}       %%%%%%%%%%%%
%%%%%%%%%%%%%%%%%%%%%%%%%%%%%%%%%
Study of the second-order coherence is useful in characterizing various quantum states of a system. For example, a system in a thermal state has a Lorentzian $g^{(2)}(\tau)$ with $1 \leqslant g^{(2)}(\tau)\leqslant 2$, and that in a coherent state has a constant $g^{(2)}(\tau)=1$ indicating the presence of lasing \cite{Kuusela}. Finally, $0 \leqslant g^{(2)}(\tau)< 1$ represents a system in a non-classical state \cite{Loudon}. Hence, to characterize the coupled-mode levitated system, we also study the second-order correlation function for it. The second-order correlation function can be expressed in terms of the creation and annihilation operators as,
\begin{eqnarray}
g^{(2)}_{j}(\tau)&=&\frac{\langle \hat{a}^{\dag}_{j}(t)\hat{a}^{\dag}_{j}(t+\tau)\hat{a}_{j}(t+\tau)\hat{a}_{j}(t)\rangle}{\langle a^{\dag}_{j}(t)\hat{a}_{j}(t)\rangle^{2}}
\label{definition correlation function},
\end{eqnarray}
where $\hat{a}^{\dag}_{j}=(Q_{j}-iP_{j})/2$, and $\hat{a}_{j}=(Q_{j}+iP_{j})$, with $j \in \{x,y\}$. Further, $\langle \rangle$ indicates an ensemble average and $\tau$ represents time delay.
%%%%%%%%%%%%%%%%%%%%%%%%%%%%%%%%%%%%%%%%%%%%%%%%%%%%%%%%%%%%%
%%%%%%%%%%%%%%%%%%%%%%%%%%%%%%%%%%%%%%%%%%%%%%%%%%%%%%%%%%%%%\textcolor{red}{\cite{Kuusela}}
\section{Results and Discussion}    
\label{RD}         %%%%%%%%%%%%%%%%%%%%%%%%
%%%%%%%%%%%%%%%%%%%%%%%%%%%%%%%%%%%%%%%%%
\subsection{PT symmetry}
In this subsection, we will theoretically demonstrate the existence of PT symmetry in a coupled-mode levitated nanoparticle. With this aim in mind, we first study the behavior of the eigenvalues of the Hamiltonian of Eqs.~(\ref{Eq_3bc}). The eigenvalues of the Hamiltonian are 
\begin{equation}
\lambda_{\pm}=-i\Gamma\pm\sqrt{-\Gamma^{2}+4(\beta^{2}+\gamma_{gx}(\gamma_{gy}-\gamma_{ay}))},
\end{equation} 
where $\Gamma=\gamma_{gx}+(\gamma_{gy}-\gamma_{ay}))$, and $\beta=\sqrt{\beta_{x}\beta_{x}}$. The variation of these eigenvalues with coupling strength is presented in Fig~\ref{fig:fig2}. It can be seen from Fig~\ref{fig:fig2} that, for the parametric regime where $\beta > \gamma $ with $\gamma_{gx}=\gamma_{gy}=\gamma$, and $\gamma_{ay}=2\gamma$, the imaginary part of both the eigenvalues are zero, while the real parts are non-zero depicting a PT-Symmetric behavior \cite{Xu}. 
	
Further, to have a better understanding of this PT-symmetric behavior, we study the dynamics of oscillator {\it x} and {\it y} modes, with {\it x} mode acting as a linearly damped oscillator and {\it y} mode as an oscillator with linear gain, using Eqs.~(\ref{Eq_2ab})-(\ref{Eq_3b}). At first, we probe the dynamics of the oscillator modes in the uncoupled state. Figures~\ref{fig:fig3a} and ~\ref{fig:fig3c} depict the behavior of oscillator displacements for the {\it x} and {\it y} mode, respectively when the modes are uncoupled. It can be seen that, in their uncoupled state the amplitude of oscillation for the {\it x}-mode decreases depicting a damped oscillator while for the {\it y}-mode it increases representing an oscillator with gain, in accordance with the initial configuration.

Next, to observe PT symmetry in the system, we now couple the {\it x} and {\it y} modes and consider equal gain and loss values, which can be arranged experimentally. In addition to this, we also consider the parametric regime where  $\beta > \gamma $. As discussed above, this parameter regime ensures that the Hamiltonian of the system has real eigenvalues \cite{Xu}. Under these conditions, as expected, the coupled system shows PT-symmetric behavior resulting in a periodic (Rabi) oscillation of mean position for both the oscillator modes with constant amplitude as depicted in Fig~\ref{fig:fig3b} and Fig~\ref{fig:fig3d}. Following our demonstration, PT symmetry in this levitated system can be utilized to further explore phenomena such as squeezing enhancement \cite{Naikoo}, efficient entanglement \cite{Ghamsari}, generation of nonclassical states \cite{Perina}, PT-symmetric phonon laser \cite{Jing}, and so on.
%%%%%%%%%%%%%%%%%%%%%%%%%%%%%%%%%%%%%%%%%%%%%%%%%%%%%%%%%%%%%%%%%%% $\gamma_{g}=0.1\omega_{0}$,  $\gamma_{g}=5\omega_{0}$,  $\gamma_{g}=0.1\omega_{0}$, and kerr parameter $U = 0.2\omega_{0}$,  $\gamma_{g}=5\omega_{0}$, and kerr
%%%%%%%%%%%%%%%%%%%%%%%%%%%%%%%%%%%%%%%%%%%%%%%%%%%%%%%%%%%%%%%%%%%%%%%%%%%%%
%         FIGURE-5                                                                                                                                                                                      
%%%%%%%%%%%%%%%%%%%%%%%%%%%%%%%%%%%%%%%%%%%%%%%%%%%%%%%%%%%%%%%%%%%%
\begin{figure}[t!]
 \centering
 \subfigure[][]{%
\label{fig:fig5a}%
\includegraphics[height=3.9cm, width=3.9cm]{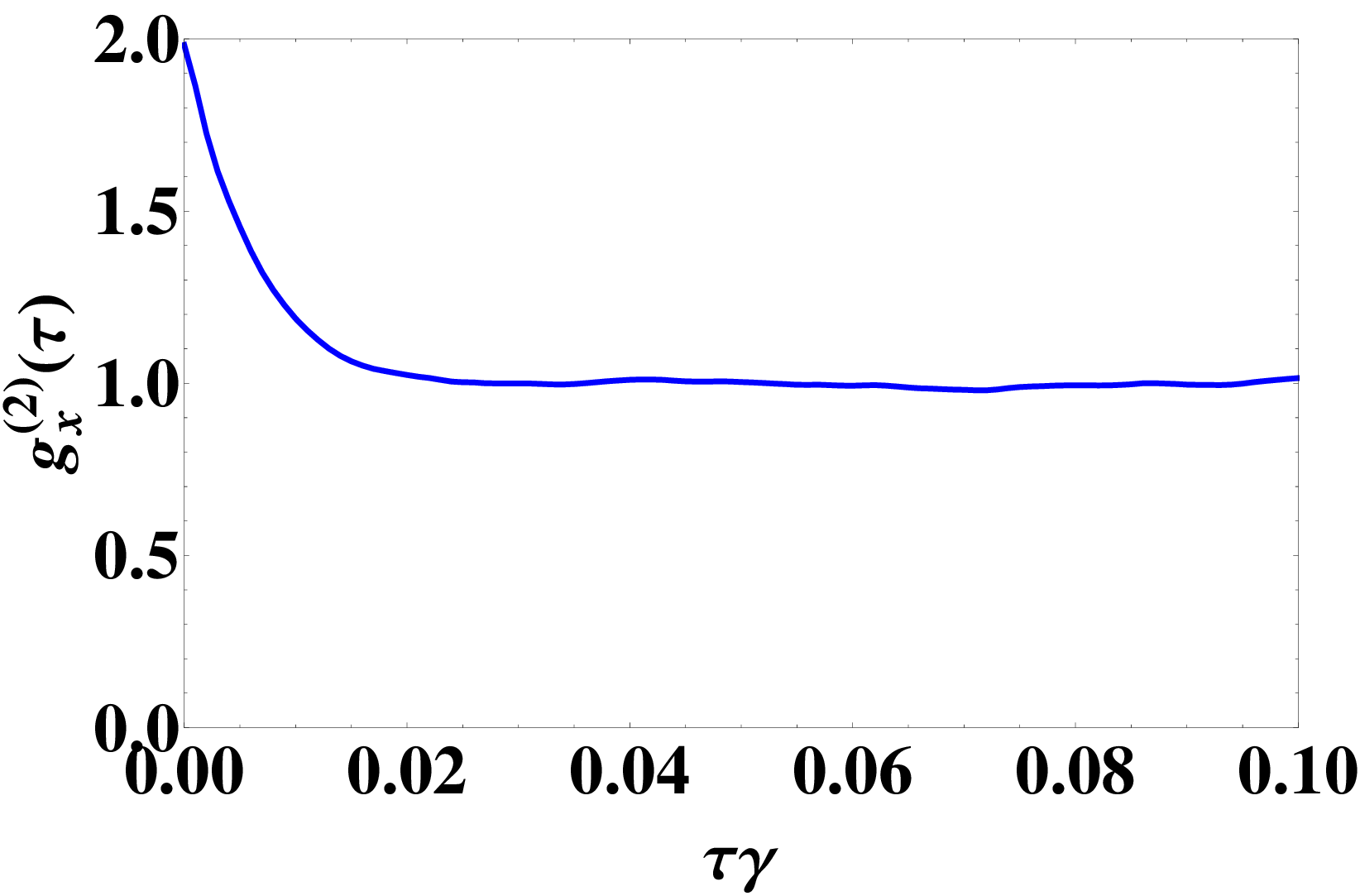}}%
\hspace{8pt}%
\subfigure[][]{%
\label{fig:fig5b}%
\includegraphics[height=3.9cm, width=3.9cm]{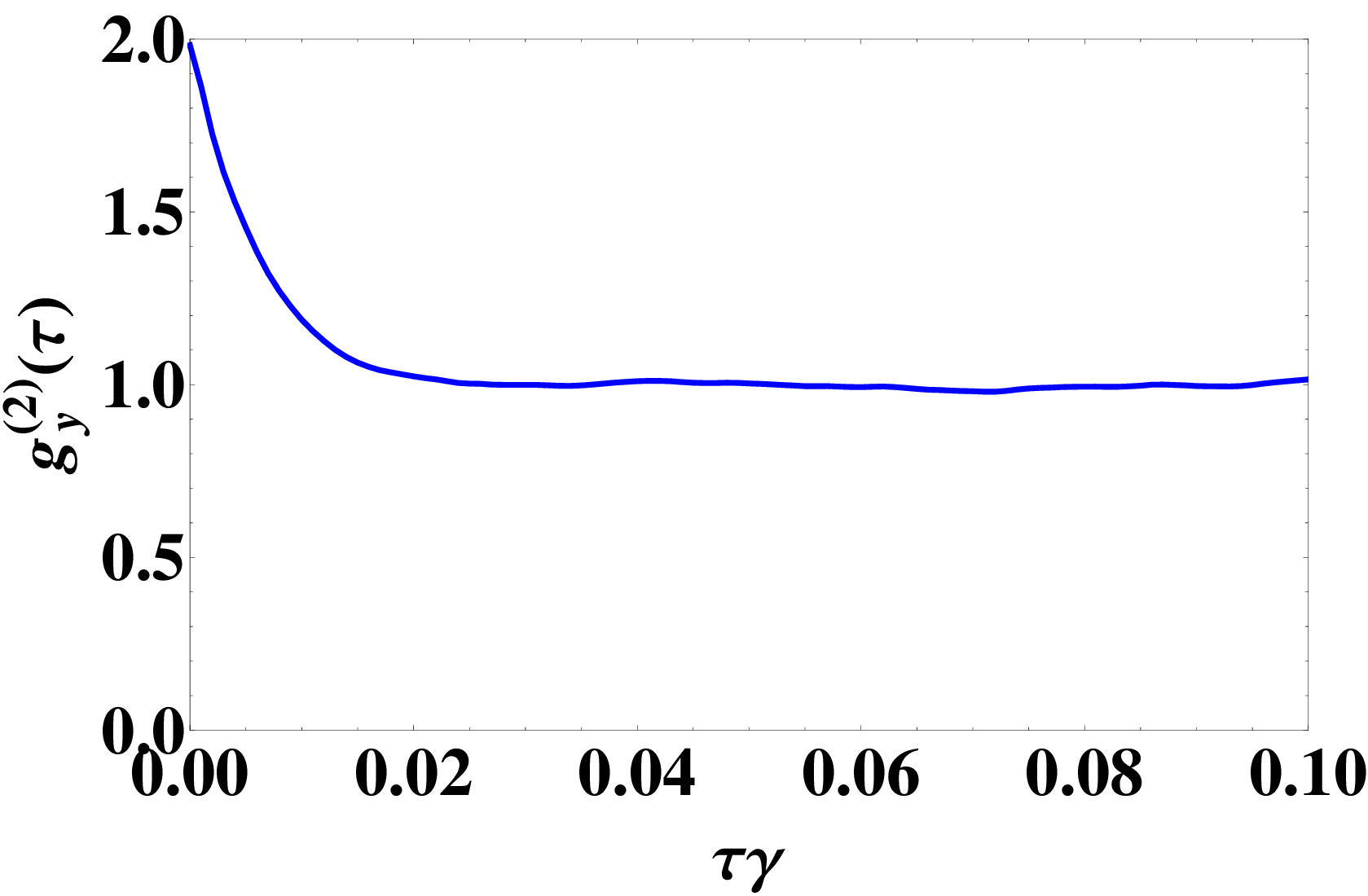}}
 \caption[]{Variation of second-order coherence with scaled time delay for the PT-symmetric system. Panels (a) and (b) show $g^{(2)}_{x}(\tau)$ for the $x$-mode and $g^{(2)}_{y}(\tau)$ for the $y$-mode, respectively. Other parameters are same as in Fig~\ref{fig:fig3}.}
 \label{fig:fig5}
 \end{figure}
%%%%%%%%%%%%%%%%%%%%%%%%%%%%%%%%%%%%%%%%%%%%%%%%%%%%%%%%%%%%%%%%%%%%%%
%%%%%%%%%%%%%%%%%%%%%%%%%%%%%%%%%%%%%%%%%%%%%%%%%%%%%%%%%%%%%%%%%%% $\gamma_{g}=0.1\omega_{0}$,  $\gamma_{g}=5\omega_{0}$,  $\gamma_{g}=0.1\omega_{0}$, and kerr parameter $U = 0.2\omega_{0}$,  $\gamma_{g}=5\omega_{0}$, and kerr parameter $U = 0.2\omega_{0}
%         FIGURE-6                                                                                                                                                                                     
%%%%%%%%%%%%%%%%%%%%%%%%%%%%%%%%%%%%%%%%%%%%%%%%%%%%%%%%%%%%%%%%%%%%
\begin{figure}[t!]
 \centering
 \subfigure[][]{%
\label{fig:fig6a}%
\includegraphics[height=3.9cm, width=3.9cm]{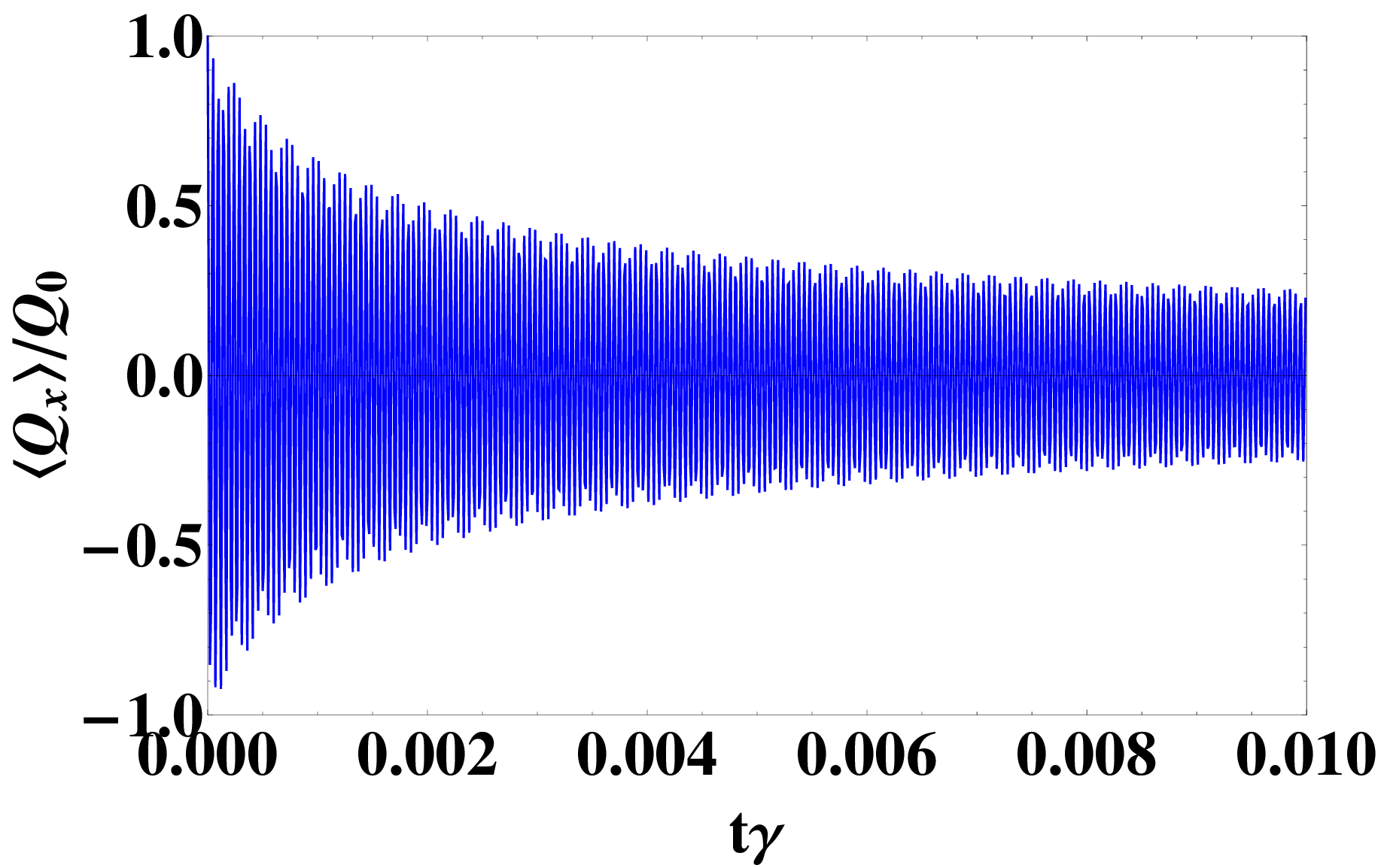}}%
\hspace{8pt}%
\subfigure[][]{%
\label{fig:fig6b}%
\includegraphics[height=3.9cm, width=3.9cm]{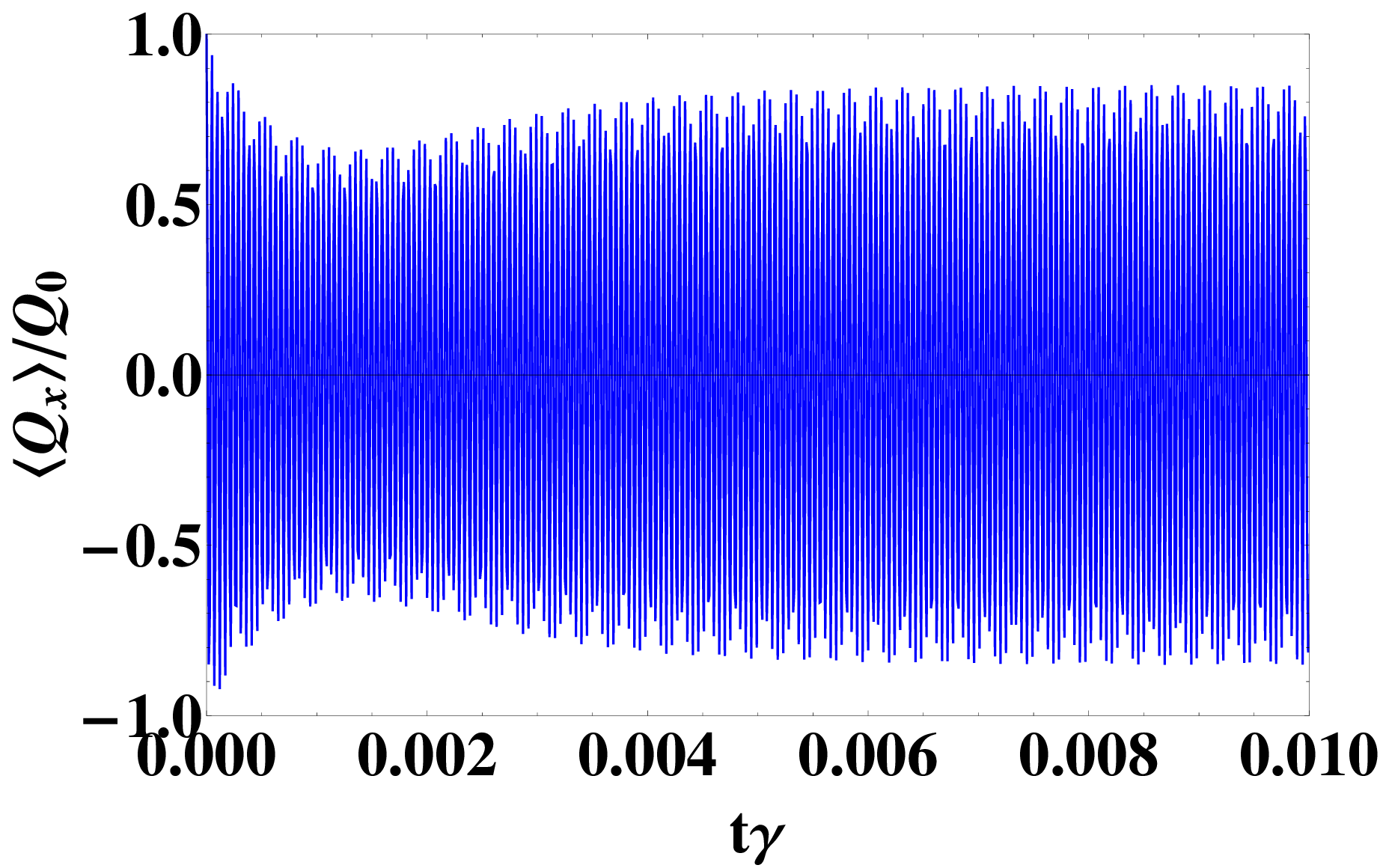}} \\
\hspace{8pt}%
\subfigure[][]{%
\label{fig:fig6c}%
\includegraphics[height=3.9cm, width=3.9cm]{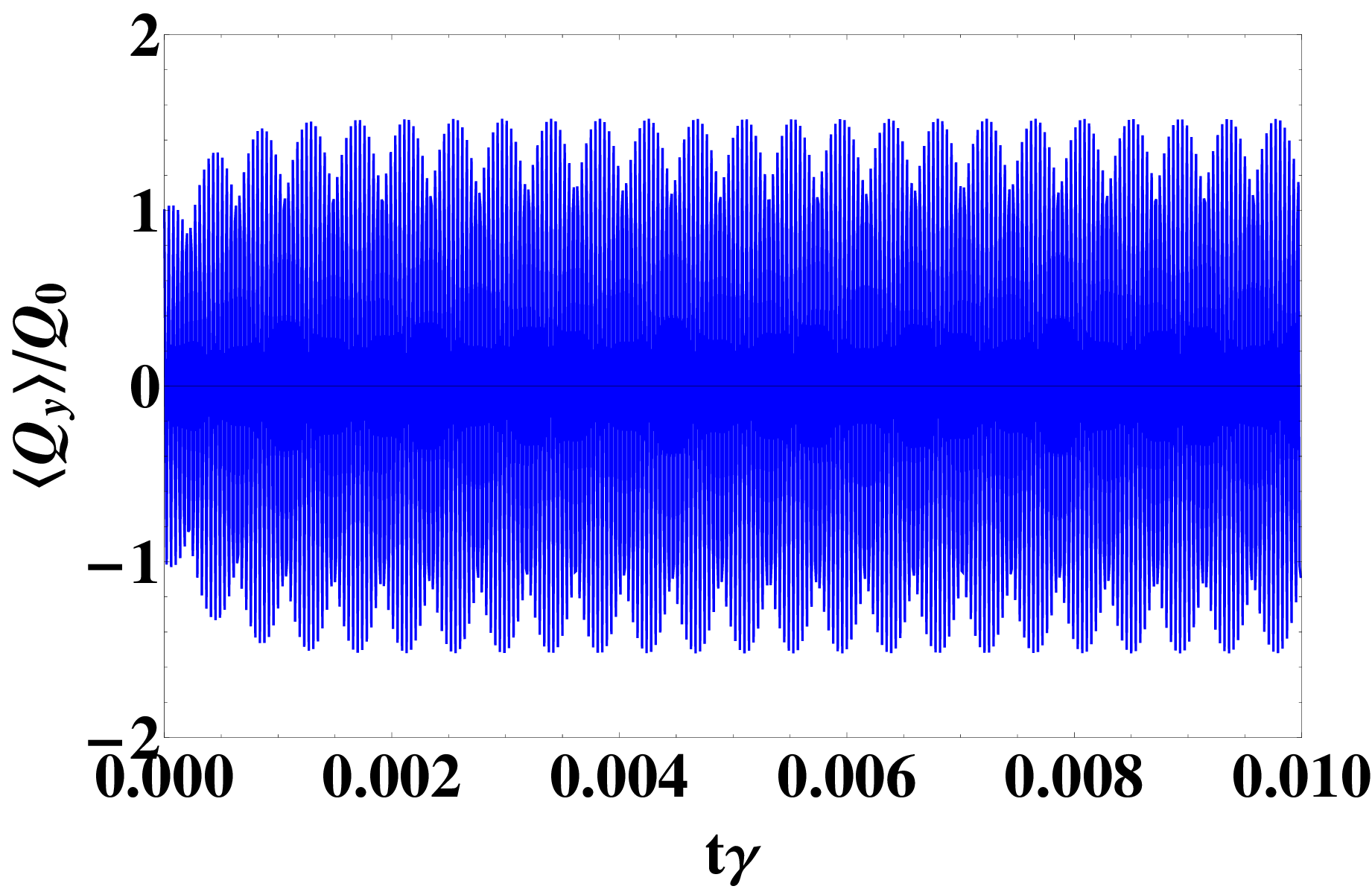}} 
\hspace{8pt}%
\subfigure[][]{%
\label{fig:fig6d}%
\includegraphics[height=3.9cm, width=3.9cm]{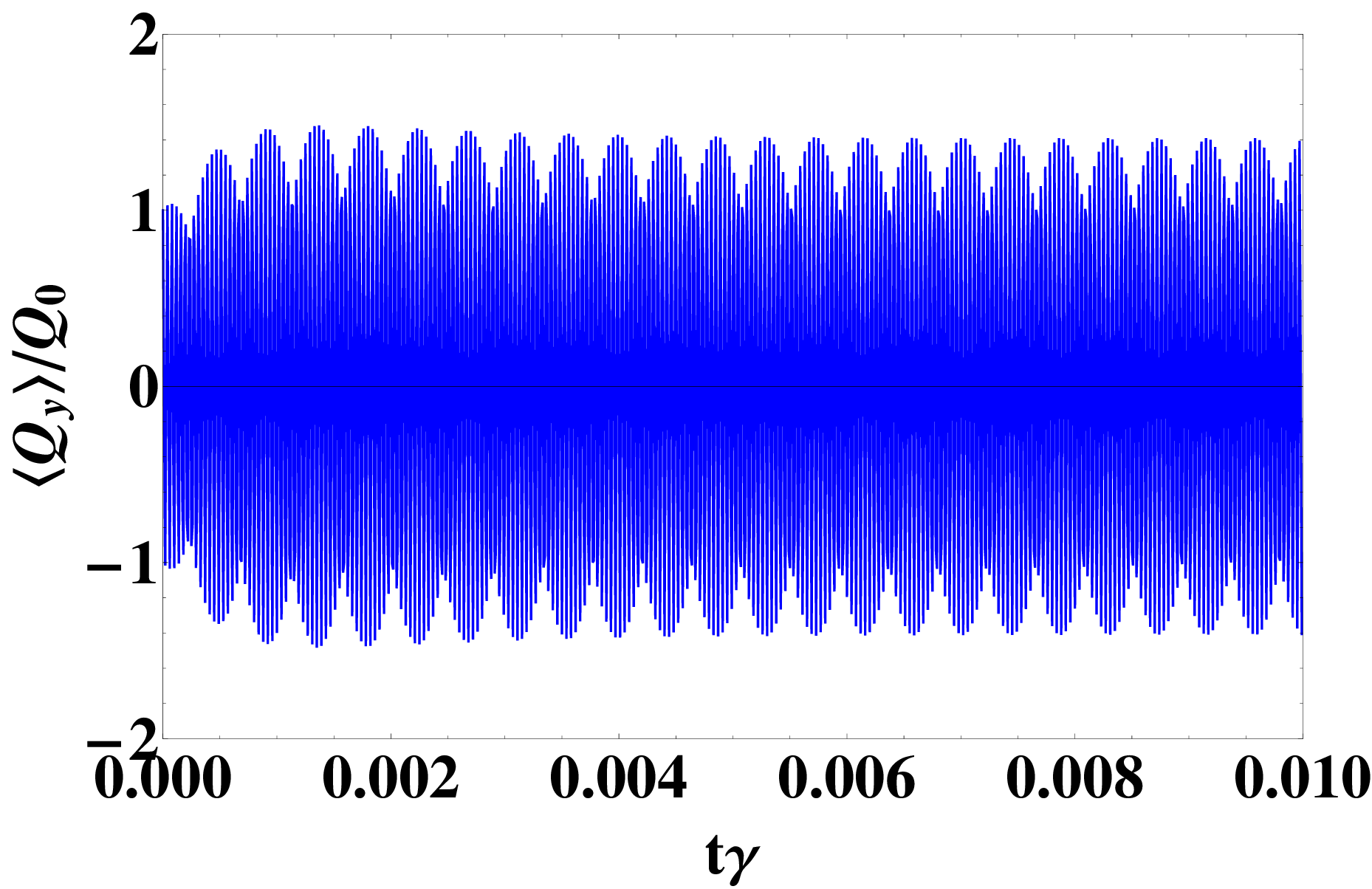}} 
 \caption[]{Induced lasing transfer in a coupled-mode levitated system. Panel (a) and (b) shows the position dynamics of $x$-mode for $\delta=0$ and  $\delta=10^{-3}$, respectively. Panel (c) and (d) shows the dynamics of $y$-mode for $\delta=0$ and  $\delta=10^{-3}$, respectively. The parameters are $\omega_{x}=130$ KHz, $\omega_{y}=160$ KHz, $\gamma_{gx}=\gamma_{gy}=\gamma=0.06$ Hz, $\gamma_{cx}=10^{-5}$ Hz, $\gamma_{cy}=10^{-5}$ Hz, $\Gamma_{cx}=10^{-6}$ Hz, $\Gamma_{cy}=10^{-6}$ Hz, $\gamma_{ay}=100$ Hz, and $Q_{0}=3\times 10^{2}$.} 
 \label{fig:fig6}
 \end{figure}
%%%%%%%%%%%%%%%%%%%%%%%%%%%%%%%%%%%%%%%%%%%%%%%%%%%%%%%%%%%%%%%%%%%%%%
We now investigate if this sustained oscillation can be linked to lasing-like behavior. For this we study the phonon dynamics as well as the second-order coherence for the coupled-mode system. Figure~\ref{fig:fig4} shows the evolution of the phonon number for both the {\it x} and {\it y} modes, both in the absence and in the presence of coupling. In the absence of coupling, the phonon population for the {\it x} mode decays exponentially while it rises for the {\it y} mode as shown in  Fig.~\ref{fig:fig4a} and Fig.~\ref{fig:fig4c}, respectively. This is simply due to the fact that in the uncoupled state, the {\it x} mode acts as a damped oscillator while {\it y} mode acts as an amplified oscillator. 

Further, when both the modes are coupled, then it is evident from  Fig.~\ref{fig:fig4b} and Fig.~\ref{fig:fig4d} that the phonon dynamics for both the modes shows oscillatory behavior. This sustained oscillation in the phonon dynamics is due to the fact that the mean phonon number $N$ is proportional to the square of mean position $Q$ ($\langle N\rangle \propto \langle Q^{2}\rangle$). Now, in the PT symmetric regime, the position dynamics of both the modes shows sustained oscillation. Hence, in this regime, the phonon dynamics for both the modes also have sustained oscillation.  

From the above study, it is apparent that the PT-symmetric system does not show a saturation-like behavior in the phonon dynamics \cite{Pettit}, which is one of the criteria for validating lasing action. 
%Moreover, even in the absence of \textcolor{blue}{photon} scattering, which can be realized by using Paul traps \cite{Bykov} and then studying the phonon dynamics starting with ground state \cite{Tebbenjohanns, Delic}, one couldn't achieve saturation like behavior in the system. In this case, the phonon population for %both the modes shows oscillation along with exponential rise \cite{Xu}. 
Hence, the self sustained oscillation of the PT-symmetric system cannot be linked to lasing-like phenomena.

In order to further verify our conclusion, we also study the second-order coherence for the PT-symmetric system, the result of which is shown in Fig.~\ref{fig:fig5}. It is evident from Fig.~\ref{fig:fig5} that $g^{(2)}_{x}(\tau)$ and $g^{(2)}_{y}(\tau)$ for both the modes show a Lorentzian-type feature indicating that both modes are in the thermal state \cite{Arkhipov}. Hence from both the studies of phonon dynamics as well as the second order coherence, it appears that lasing-like phenomena in this PT-symmetric system remains inaccessible. In the next section we turn to a configuration where one of the modes is lasing \cite{Pettit}. In such a configuration, it may be possible to attain lasing in both the modes, similar to coherent quantum state transfer phenomena where one can transfer quantum state from one mode to the other \cite{Hammerer}. 
%%%%%%%%%%%%%%%%%%%%%%%%%%%%%%%%%%%%%%%%%%%%%%%%%%%%%%%%%%%%%%%%%%%%%%%%%
\subsection{Lasing transfer}
In this subsection, we demonstrate coupling-induced lasing transfer between the two transverse modes of the levitated nanoparticle. For this, we initially consider the {\it x} mode as a linearly damped oscillator and {\it y} mode as a phonon laser which requires both amplification and cooling \cite{Pettit}. At first, we study the position dynamics for both the modes of the coupled system, the result of which is shown in Fig.~\ref{fig:fig6}. Specifically, figures~\ref{fig:fig6a} and~\ref{fig:fig6c} show the amplitudes of the $x$ and $y$ modes, respectively, of the system when both the modes are uncoupled. As expected, the $x$ mode simply damps to a lower amplitude, while the $y$ mode shows the
previously observed tangent hyperbolic rise \cite{Pettit}. However, when the modes are coupled, the {\it y} mode induces a sustained oscillation in the {\it x} mode (which was initially a damped oscillator) as shown in Fig.~\ref{fig:fig6b}.
%%%%%%%%%%%%%%%%%%%%%%%%%%%%%%%%%%%%%%%%%%%%%%%%%%%%%%%%%%%%%%%%%%% $\gamma_{g}=0.1\omega_{0}$,  $\gamma_{g}=5\omega_{0}$,  $\gamma_{g}=0.1\omega_{0}$, and kerr parameter $U = 0.2\omega_{0}$,  $\gamma_{g}=5\omega_{0}$, and kerr parameter $U = 0.2\omega_{0}
%         FIGURE-7                                                                                                                                                                                     
%%%%%%%%%%%%%%%%%%%%%%%%%%%%%%%%%%%%%%%%%%%%%%%%%%%%%%%%%%%%%%%%%%%%
\begin{figure}[t!]
 \centering
 \subfigure[][]{%
\label{fig:fig7a}%
\includegraphics[height=3.9cm, width=3.9cm]{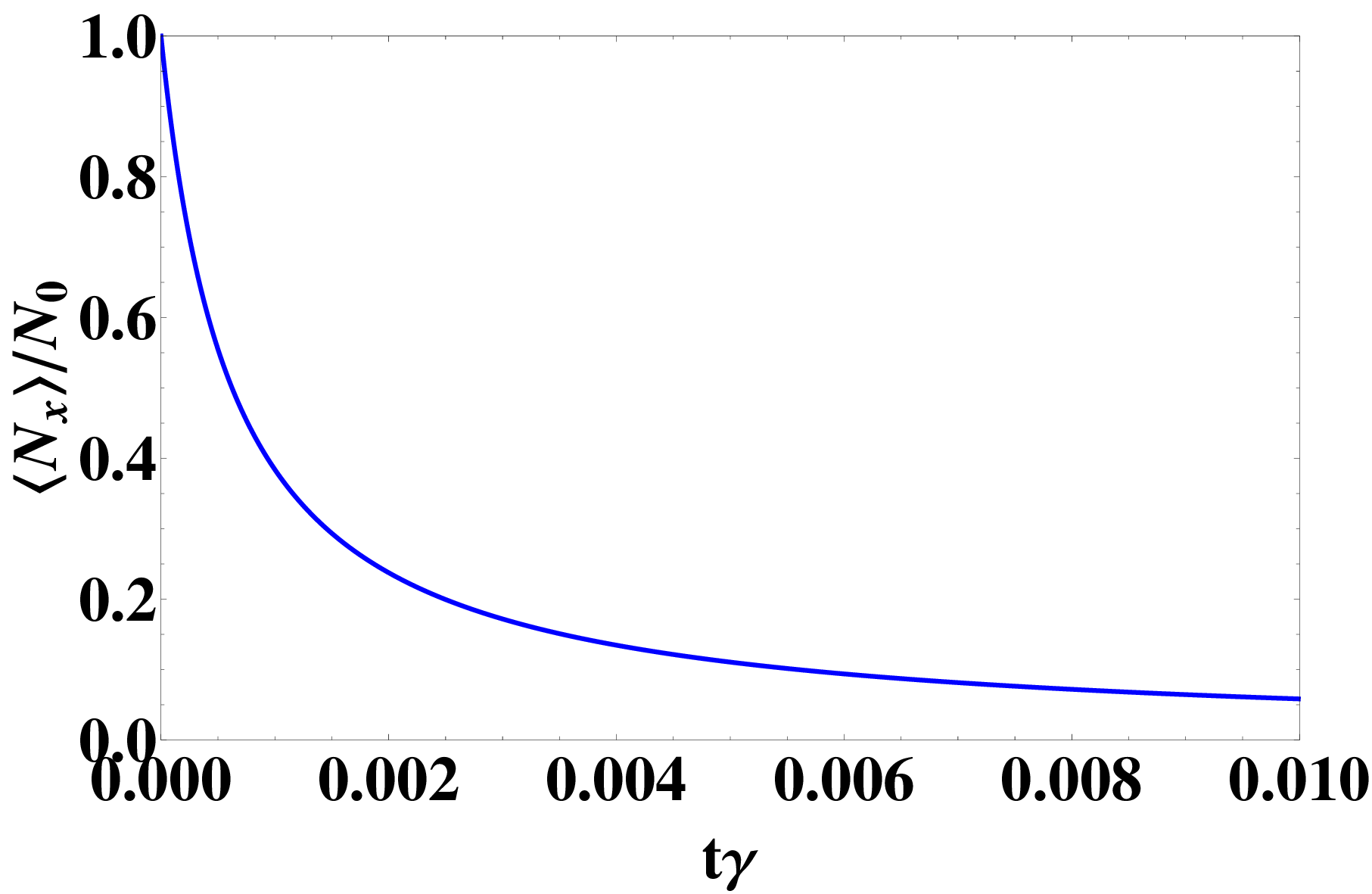}}%
\hspace{8pt}%
\subfigure[][]{%
\label{fig:fig7b}%
\includegraphics[height=3.9cm, width=3.9cm]{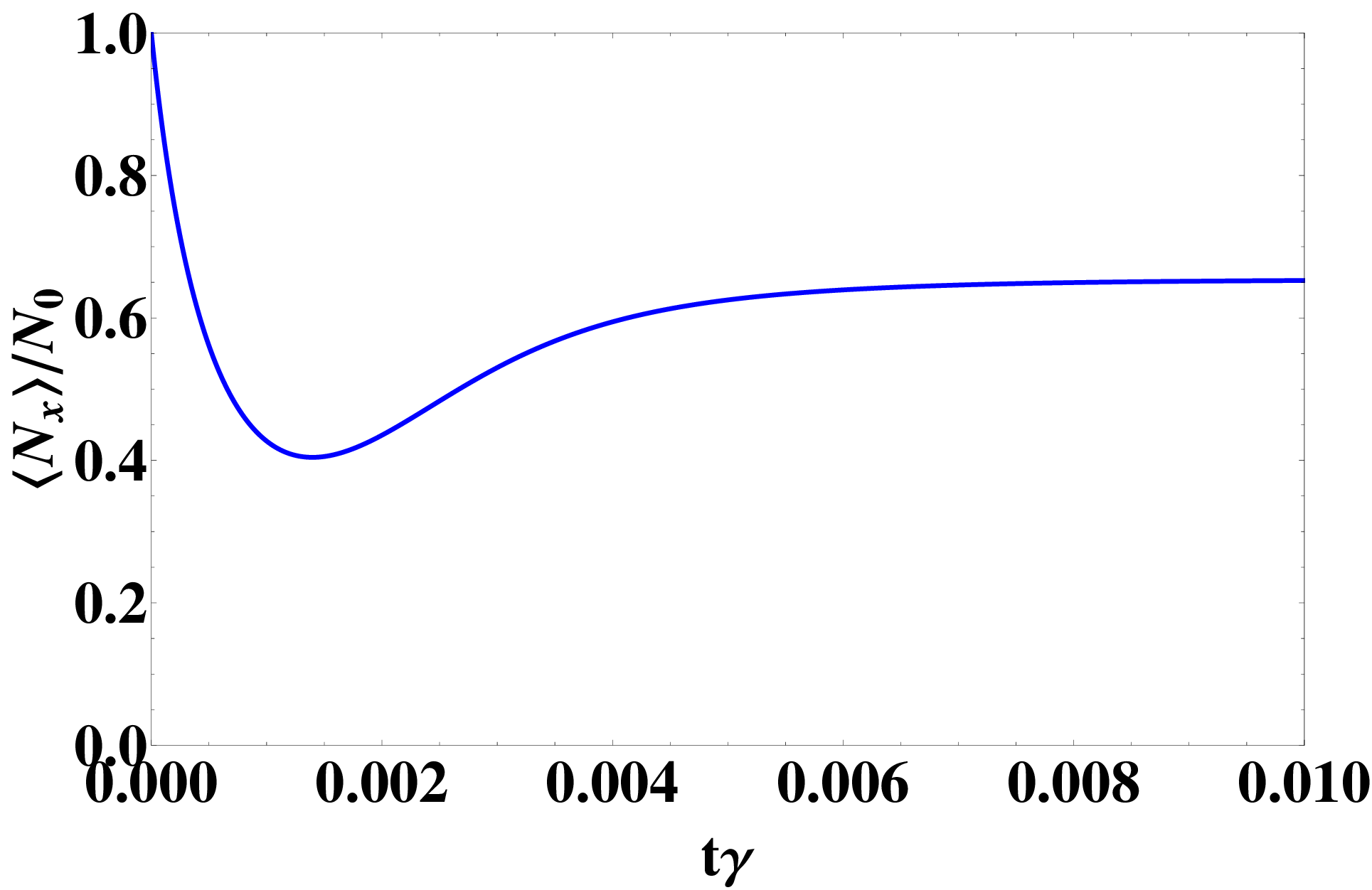}} \\
\hspace{8pt}%
\subfigure[][]{%
\label{fig:fig7c}%
\includegraphics[height=3.9cm, width=3.9cm]{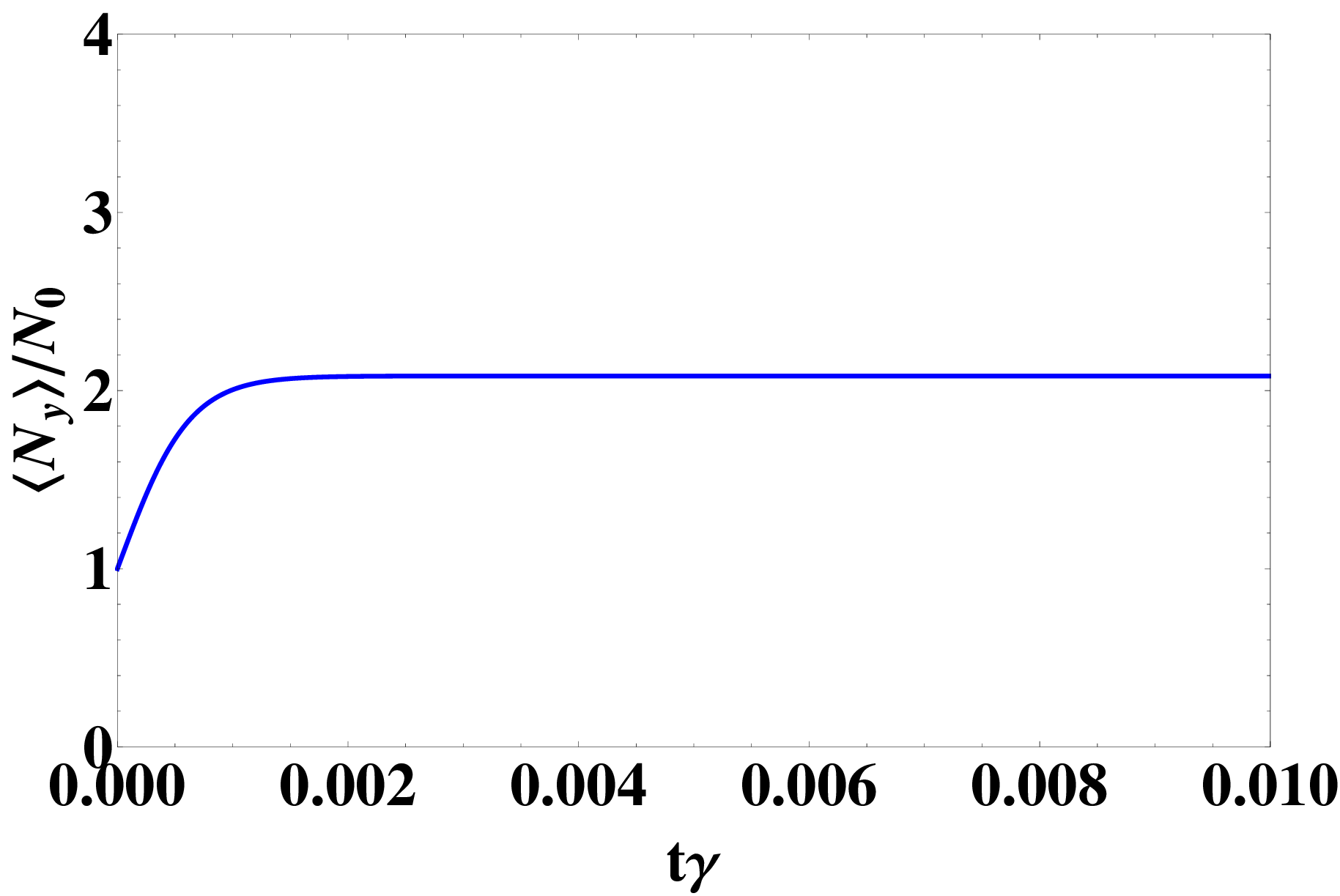}} 
\hspace{8pt}%
\subfigure[][]{%
\label{fig:fig7d}%
\includegraphics[height=3.9cm, width=3.9cm]{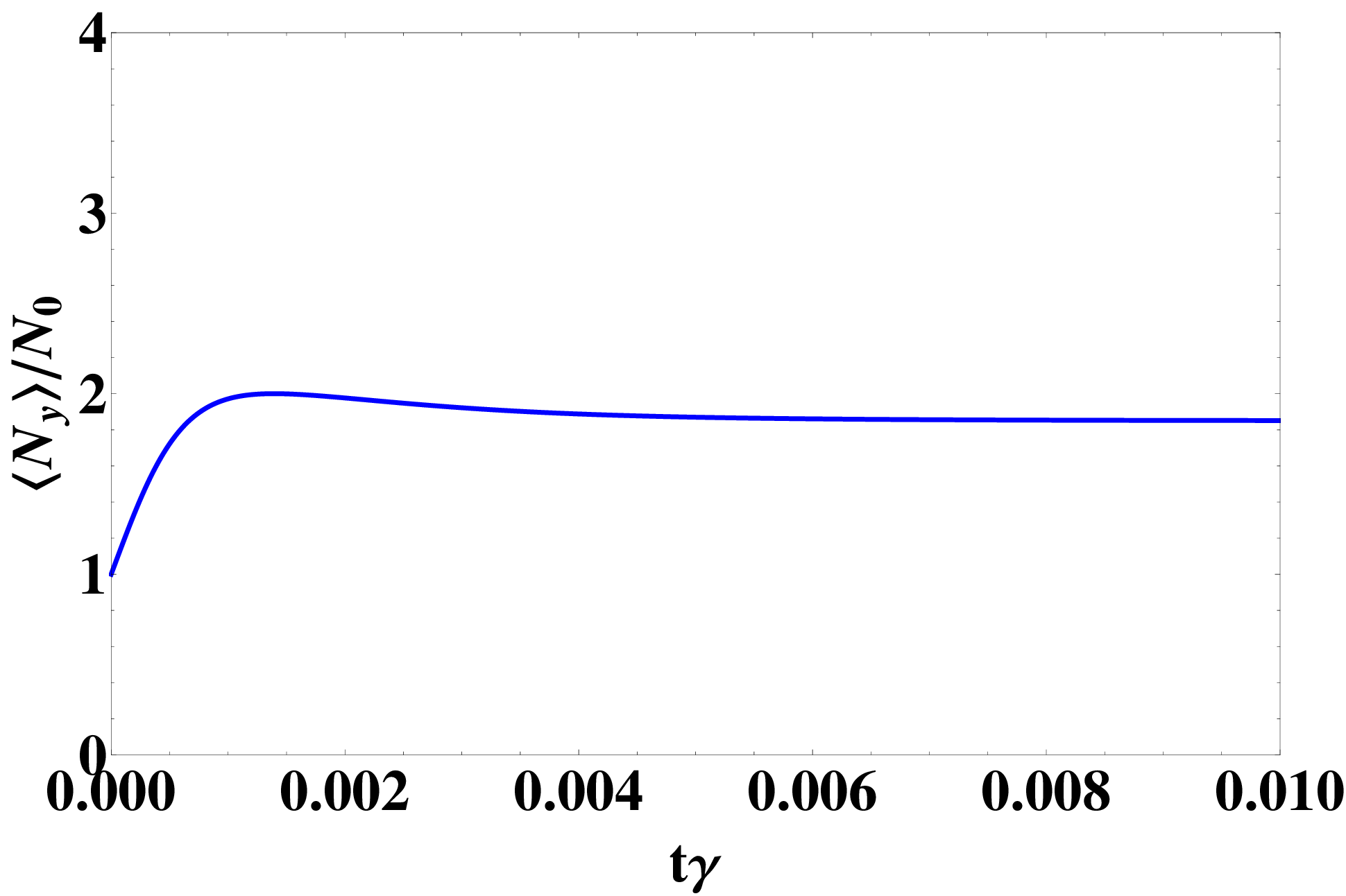}} 
 \caption[]{Phonon dynamics of a coupled-mode levitated nanoparticle. Panels (a) and (b) show the evolution of normalized phonon population of the $x$-mode for $\delta=0$ and  $\delta=10^{-3}$, respectively. Panels (c) and (d) show the evolution of phonon population of the $y$-mode for $\delta=0$ and  $\delta=10^{-3}$, respectively.  Here $D_{tx}=1$ KHz, $D_{ty}=1$ KHz, and $N_{0}=10^{5}$, while other parameters are the same as in Fig~\ref{fig:fig6}.}
 \label{fig:fig7}
 \end{figure}
%%%%%%%%%%%%%%%%%%%%%%%%%%%%%%%%%%%%%%%%%%%%%%%%%%%%%%%%%%%%%%%%%%%%%%

Further, to check that the sustained oscillation of the {\it x} mode indeed leads to lasing of the mode, we study the phonon dynamics for the coupled system. Figure~\ref{fig:fig7} shows the dynamical evolution of the phonon population in {\it x} and {\it y} modes both in the absence and in the presence of coupling. 
%\textcolor{red}{As expected with the initial conditions, the phonon population in the {\it x} mode follows a tangent hyperbolic like decay \cite{Rodenburg}} (\textcolor{red}{Comment:- Are you sure it decreases exponentially? Send me a plot on a logscale. In the phonon laser the variation is not exponential it is tangent hyperbolic. Response:- Yes, you are correct. The following line is added above {\it As expected with the initial conditions, the phonon population in the {\it x} mode follows a tangent hyperbolic like decay \cite{Rodenburg}.}), 
In the absence of coupling, as expected with the initial conditions, the phonon population in the {\it x} mode follows an  exponential decay, while for the {\it y} mode, the phonon population shows a saturation effect since it is  a lasing mode, as shown in  Fig.~\ref{fig:fig7a} and  Fig.~\ref{fig:fig7c}. With the coupling in effect, it can be seen from Fig.~\ref{fig:fig7b} that the phonon population in {\it x} mode attains a saturation value, which points to a lasing-like phenomenon. Further, the phonon population in the {\it y} mode slightly decreases while preserving the saturation effect, as seen in Fig.~\ref{fig:fig7d}. This decrease in the phonon population is due to the coupling of {\it y} mode to the damped {\it x} mode, which induces extra damping in the {\it y} mode.
%%%%%%%%%%%%%%%%%%%%%%%%%%%%%%%%%%%%%%%%%%%%%%%%%%%%%%%%%%%%%%%%%%% $\gamma_{g}=0.1\omega_{0}$,  $\gamma_{g}=5\omega_{0}$,  $\gamma_{g}=0.1\omega_{0}$, and kerr parameter $U = 0.2\omega_{0}$,  $\gamma_{g}=5\omega_{0}$, and kerr parameter $U = 0.2\omega_{0}
%         FIGURE-8                                                                                                                                                                                     
%%%%%%%%%%%%%%%%%%%%%%%%%%%%%%%%%%%%%%%%%%%%%%%%%%%%%%%%%%%%%%%%%%%%
\begin{figure}[t!]
 \centering
 \subfigure[][]{%
\label{fig:fig8a}%
\includegraphics[height=3.9cm, width=3.9cm]{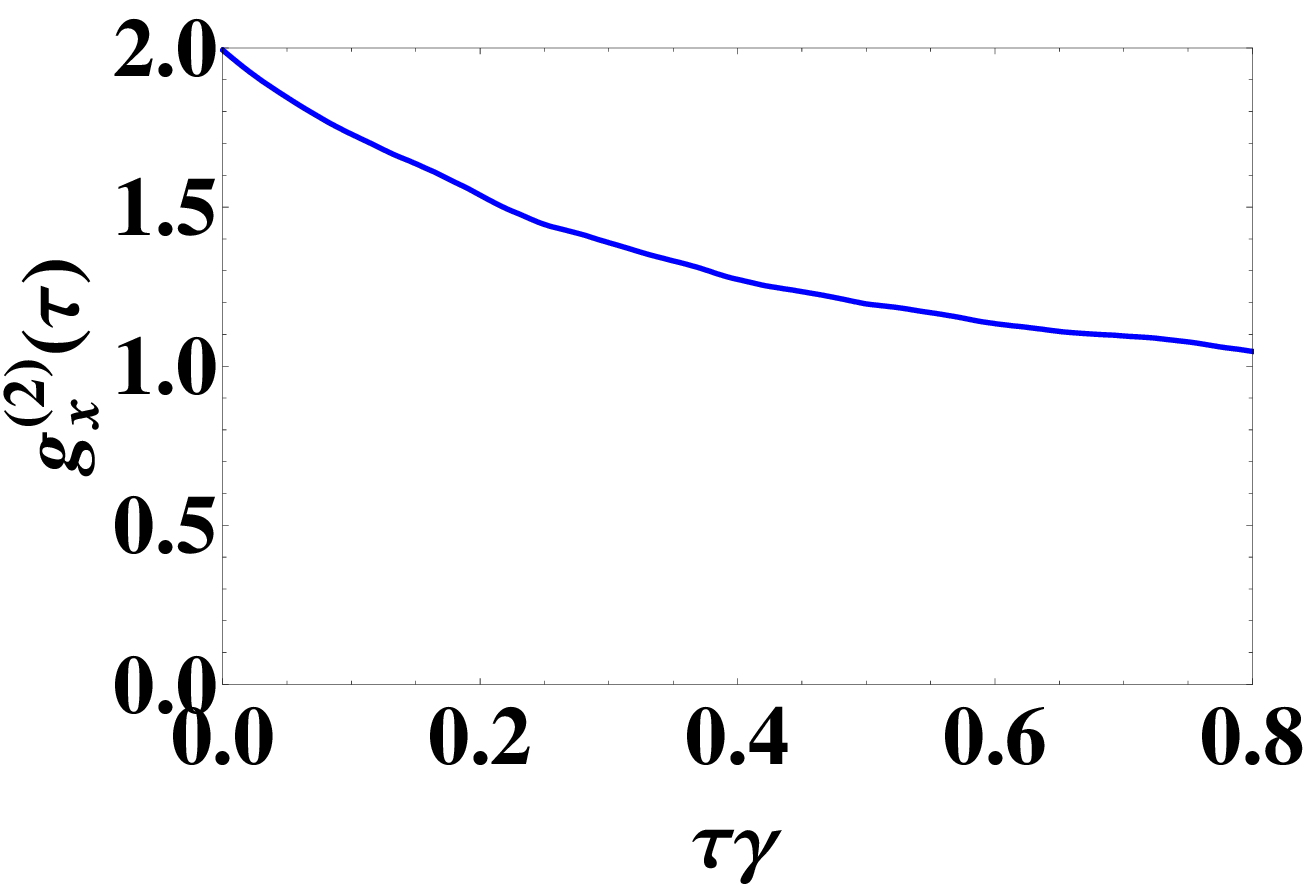}}%
\hspace{8pt}%
\subfigure[][]{%
\label{fig:fig8b}%
\includegraphics[height=3.9cm, width=3.9cm]{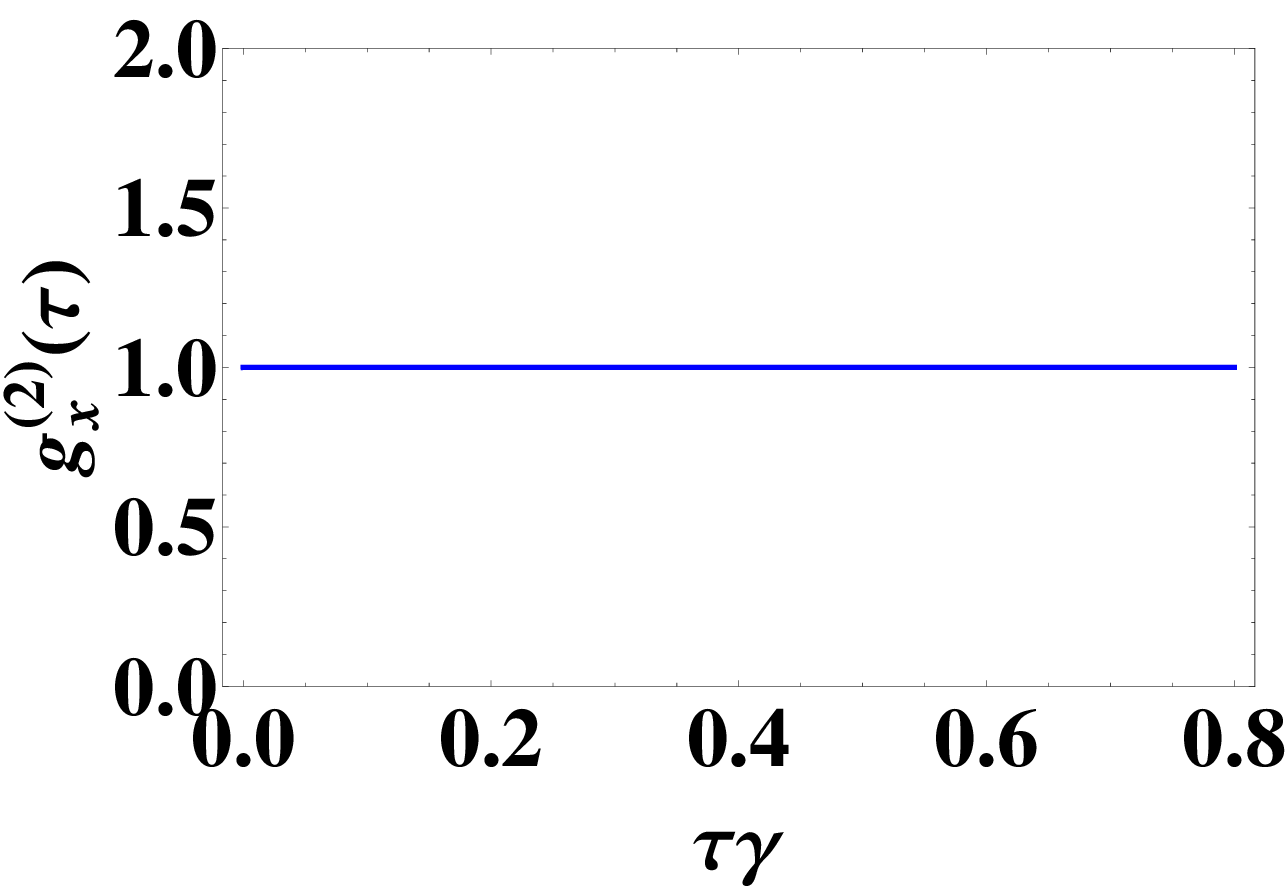}} \\
\hspace{8pt}%
\subfigure[][]{%
\label{fig:fig8c}%
\includegraphics[height=3.9cm, width=3.9cm]{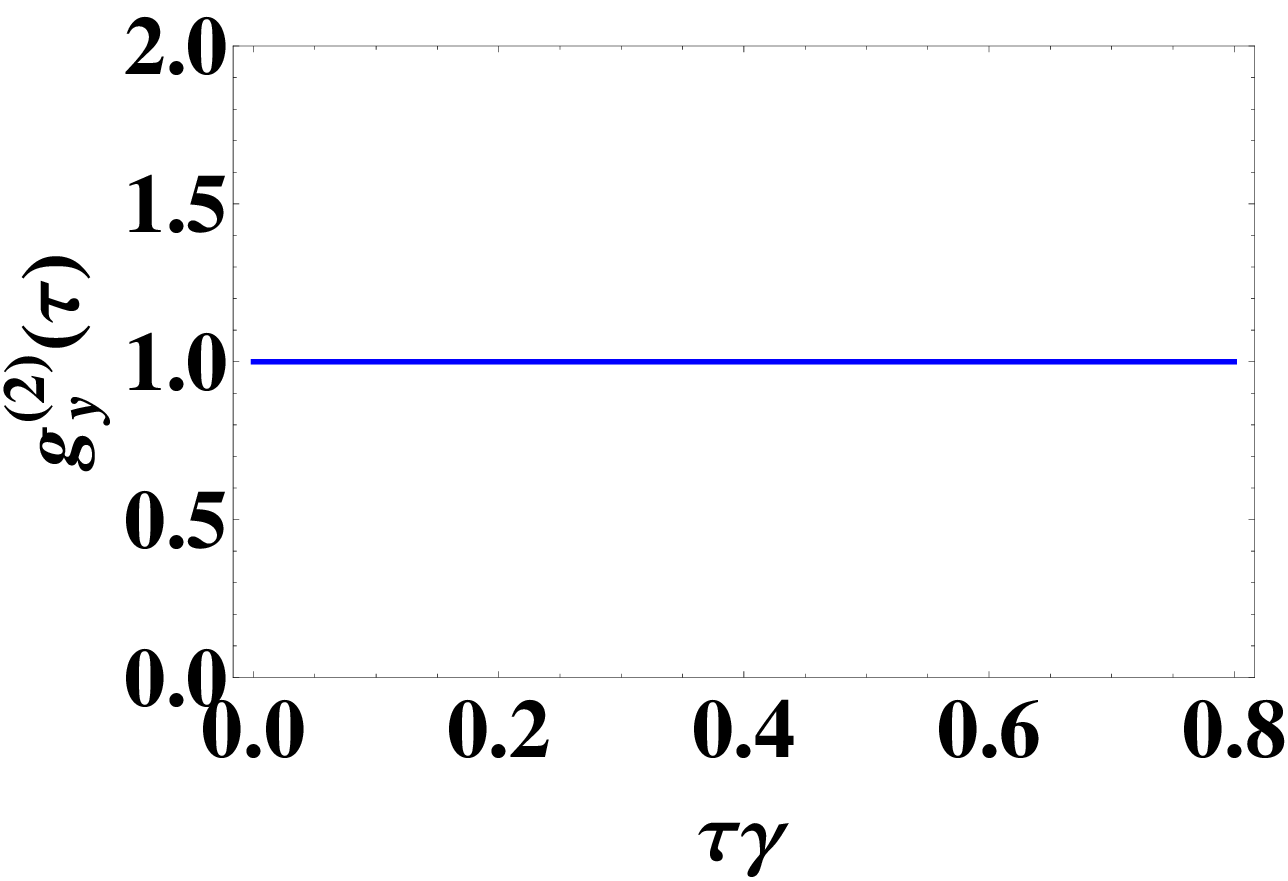}} 
\hspace{8pt}%
\subfigure[][]{%
\label{fig:fig8d}%
\includegraphics[height=3.9cm, width=3.9cm]{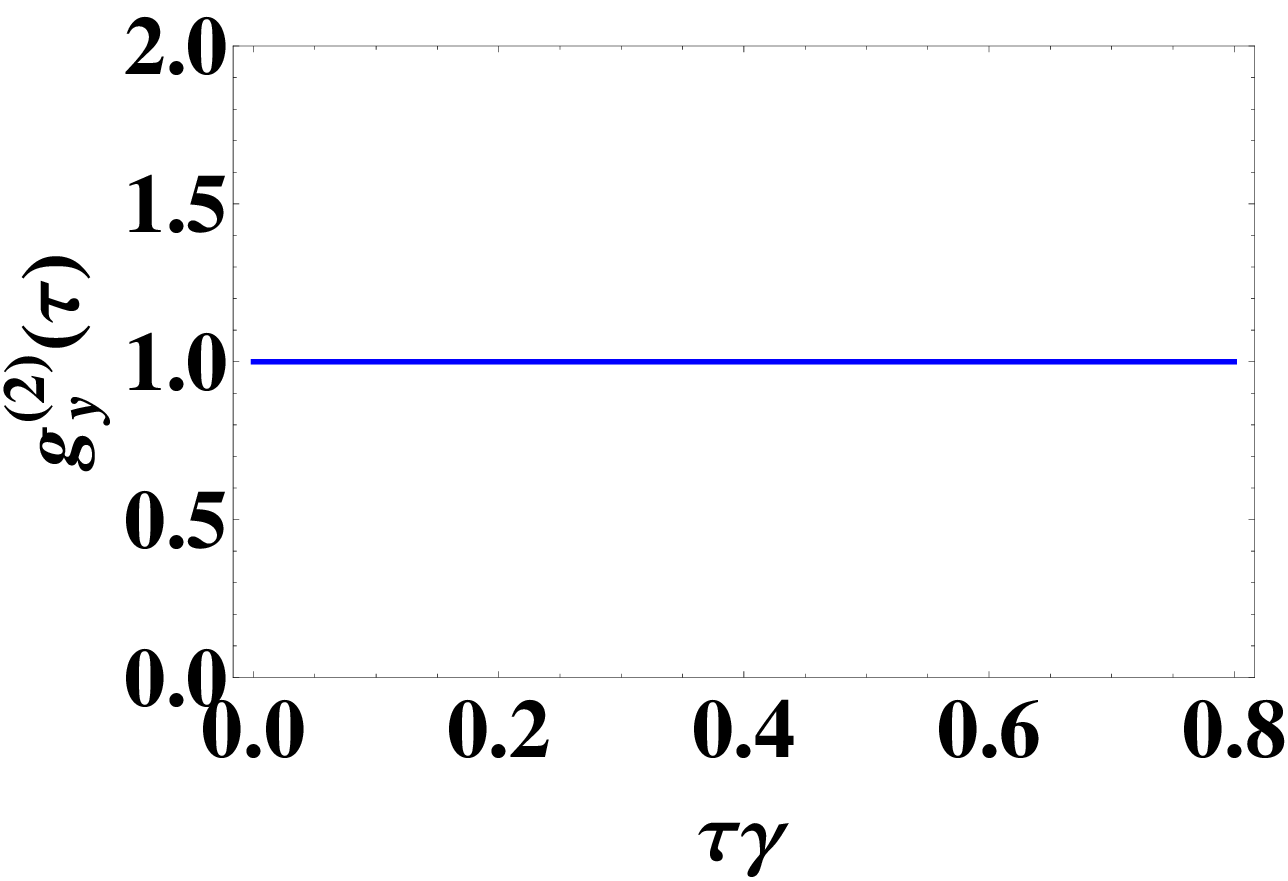}} 
 \caption[]{The second-order coherence as a function of scaled time delay for the coupled-mode levitated system. Panels (a) and (b) show $g^{(2)}_{x}(\tau)$ of the $x$-mode for $\delta=0$ and  $\delta=10^{-3}$, respectively. Panels (c) and (d) show $g^{(2)}_{y}(\tau)$ of the $y$-mode for $\delta=0$ and  $\delta=10^{-3}$, respectively. Other parameters are same as in Fig~\ref{fig:fig6}.}
 \label{fig:fig8}
 \end{figure}
%%%%%%%%%%%%%%%%%%%%%%%%%%%%%%%%%%%%%%%%%%%%%%%%%%%%%%%%%%%%%%%%%%%%%%

Next, we also study the second-order coherence for the coupled system to fully establish induced lasing transfer in the system. It can be seen from Fig.~\ref{fig:fig8a} and Fig.~\ref{fig:fig8c} that initially, $g^{(2)}_{x}(\tau)$ for the {\it x} mode has a Lorentzian profile depicting the mode to be in a thermal state and $g^{(2)}_{y}(\tau)$ for the {\it y} mode has a constant profile representing it to be in a coherent state, respectively  \cite{Kuusela}. However, with the introduction of coupling in the system, $g^{(2)}_{x}(\tau)$ for the {\it x} mode also shows a constant profile, indicating that the mode has evolved to a coherent state, while the {\it y} mode remains as before, as shown in  Fig.~\ref{fig:fig8b} and Fig.~\ref{fig:fig8d}, respectively. Hence, our study of both phonon dynamics and the second order coherence confirms that the induced lasing transfers between the modes in the coupled system. 

These results can stimulate studies on outcoupling of phonon lasers and obtaining high fidelity quantum state transfer \cite{Hammerer} as well as efficient synchronization \cite{Roulet,Stannigel}.
%%%%%%%%%%%%%%%%%%%%%%%%%%%%%%%%%%%%%%%%%%%%%%%%%%%%%%%%%%%%%%%%%%%%%%%%
%#################################################################%
\subsection{Force sensing}
In this section, we study force sensing using a coupled-mode levitated nanoparticle. 
In order to calculate the force sensitivity, we first find the optomechanical susceptibilities by taking the derivative of the equations Eqs.~(\ref{Eq_2a}-\ref{Eq_2b}) with respect to time; after some simplifications we find that
%%%%%%%%%%%%%%%%%%%%%%%%%%%%%%%%%%%%%%%%%%%%%%%%%%%%%%%%%%%%%%%%%%%%%%
\begin{align}
\ddot{q}_{x}&=-\omega^{2}_{x}q_{x}-\frac{\kappa \delta }{m}\cos(\omega_{r}t) q_{y} \nonumber \\
	             &-2\left[\gamma_{gx}+12\gamma_{cx}(2\langle N_{x} \rangle+1)\right]\dot{q}_{x} +\frac{F_{x}}{m},\\
\ddot{q}_{y}&=-\omega^{2}_{y}q_{y}-\frac{\kappa \delta }{m}\cos(\omega_{r}t) q_{x} \nonumber \\
	             &-2\left[\gamma_{gy}+12\gamma_{cy}(2\langle N_{y} \rangle+1)\right]\dot{q}_{y}+\frac{F_{y}}{m},
\end{align}
%%%%%%%%%%%%%%%%%%%%%%%%%%%%%%%%%%%%%%%%%%%%%%%%%%%%%%%%%%%%%%%%%%%%%%
where $F_{j}$=$F^{j}_{T}+F^{j}_{Fa}+F^{j}_{Fc}$, and $F^{j}_{T}$=$\sqrt{2K_{B}Tm\gamma_{gj}}\xi_{T}$, $F^{j}_{Fa}$=$\sqrt{D_{tj}\hbar\omega_{j}m}\xi_{Fa}$, $F^{j}_{Fc}$=$12q^{2}_{j}\sqrt{\frac{\Gamma^{2}_{cj}m^{3}\omega^{3}_{j}}{\hbar\gamma_{cj}}}\xi_{Fc}$, with $j \in \{x,y\}$.
%%%%%%%%%%%%%%%%%%%%%%%%%%%%%%%%%%%%%%%%%%%%%%%%%%%%%%%%%%%%%%%%%%%%%%%%
We now take a Fourier transform of the above equations and rewrite into the following form
%%%%%%%%%%%%%%%%%%%%%%%%%%%%%%%%%%%%%%%%%%%%%%%%%%%%%%%%%%%%%%%%%%%%%%
%\begin{align}
%-\omega^{2}q_{x}(\omega)&=-\omega^{2}_{x}Q_{x}(\omega)-\kappa \omega^{2}_{x} Q_{y}(\omega) \nonumber \\
%&-i2(\gamma_{gx}+\frac{24\gamma_{cx}m\omega_{x}}{\hbar}(2\langle N_{x} \rangle+1))\omega q_{x}(\omega) \nonumber \\
                     %&+\frac{F^{x}_{T}(\omega)}{m}+\frac{F^{x}_{Fa}(\omega)}{m}+\frac{F^{x}_{Fc}(\omega)}{m}, \\
%-\omega^{2}q_{y}(\omega)&=-\omega^{2}_{y}q_{y}(\omega)-\kappa \omega^{2}_{y} q_{x}(\omega) \nonumber \\
%&-i2(\gamma_{gy}-\gamma_{ay}+\frac{24\gamma_{cy}m\omega_{y}}{\hbar}(2\langle N_{y} \rangle+1))\omega q_{y}(\omega) \nonumber \\
                     %&+\frac{F^{y}_{T}(\omega)}{m}+\frac{F^{y}_{Fa}(\omega)}{m}+\frac{F^{y}_{Fc}(\omega)}{m}.
%\end{align}
%%%%%%%%%%%%%%%%%%%%%%%%%%%%%%%%%%%%%%%%%%%%%%%%%%%%%%%%%%%%%%%%%%%%%%
%%%%%%%%%%%%%%%%%%%%%%%%%%%%%%%%%%%%%%%%%%%%%%%%%%%%%%%%%%%%%%%%%%%
\begin{align}
\label{Eq_11}
\left[(\omega^{2}_{x}-\omega^{2})+i\omega \Gamma_{x}\right]q_{x}(\omega)+\frac{\kappa \delta }{2m} q_{y}(\omega-\omega_{r}) \nonumber \\
+\frac{\kappa \delta }{2m} q_{y}(\omega+\omega_{r})&=\frac{F_{x}(\omega)}{m},\\
\left[(\omega^{2}_{y}-\omega^{2})+i\omega \Gamma_{y}\right]q_{y}(\omega)+\frac{\kappa \delta }{2m} q_{x}(\omega-\omega_{r})\nonumber \\
+\frac{\kappa \delta }{2m} q_{x}(\omega+\omega_{r})&=\frac{F_{y}(\omega)}{m},
\label{Eq_12}
\end{align}
%%%%%%%%%%%%%%%%%%%%%%%%%%%%%%%%%%%%%%%%%%%%%%%%%%%%%%%%%%%%%%%%%%%
where $\Gamma_{x}=2\left[\gamma_{gx}+12\gamma_{cx}(2\langle N_{x} \rangle+1)\right]$, and $\Gamma_{y}=2\left[\gamma_{gy}+12\gamma_{cy}(2\langle N_{y} \rangle+1)\right]$.

Next, we solve Eq.~(\ref{Eq_11}) and Eq.~(\ref{Eq_12}) and write the solutions in the form $q_{x}(\omega)=\chi_{x}(\omega)F_{x}(\omega)$ and $q_{y}(\omega)=\chi_{y}(\omega)F_{y}(\omega)$, where $\chi_{x}(\omega)$ and $\chi_{y}(\omega)$ are the optomechanical susceptibilities  for the {\it x} and {\it y} modes, respectively. Here, we consider both the modes to be in nearly identical conditions such as subjected to the same gas damping, scattering rates, feedback rates and at the same effective temperature. Under this assumption, the mean effective forces acting on both modes can be considered to be nearly equal i.e., $\langle|F_{x}(\omega)|\rangle  \approx \langle|F_{y}(\omega)|\rangle$. In this approximation the optomechanical susceptibilities can be expressed as
%%%%%%%%%%%%%%%%%%%%%%%%%%%%%%%%%%%%%%%%%%%%%%%%%%%%%%%%%%%%%%%%%
%%%%%%%%%%%%%%%%%%%%%%%%%%%%%%%%%%%%%%%%%%%%%%%%%%%%%%%%%%%%%%%%%%%%%%
%%%%%%%%%%%%%%%%%%%%%%%%%%%%%%%%%%%%%%%%%%%%%%%%%%%%%%%%%%%%%%%%%%% $\gamma_{g}=0.1\omega_{0}$,  $\gamma_{g}=5\omega_{0}$,  $\gamma_{g}=0.1\omega_{0}$, and kerr parameter $U = 0.2\omega_{0}$,  $\gamma_{g}=5\omega_{0}$, and kerr parameter $U = 0.2\omega_{0}
%         FIGURE-9                                                                                                                                                                                       
%%%%%%%%%%%%%%%%%%%%%%%%%%%%%%%%%%%%%%%%%%%%%%%%%%%%%%%%%%%%%%%%%%%%
\begin{figure}[t!]
 \centering
 \subfigure[][]{%
\label{fig:fig9a}%
\includegraphics[height=3.8cm, width=3.9cm]{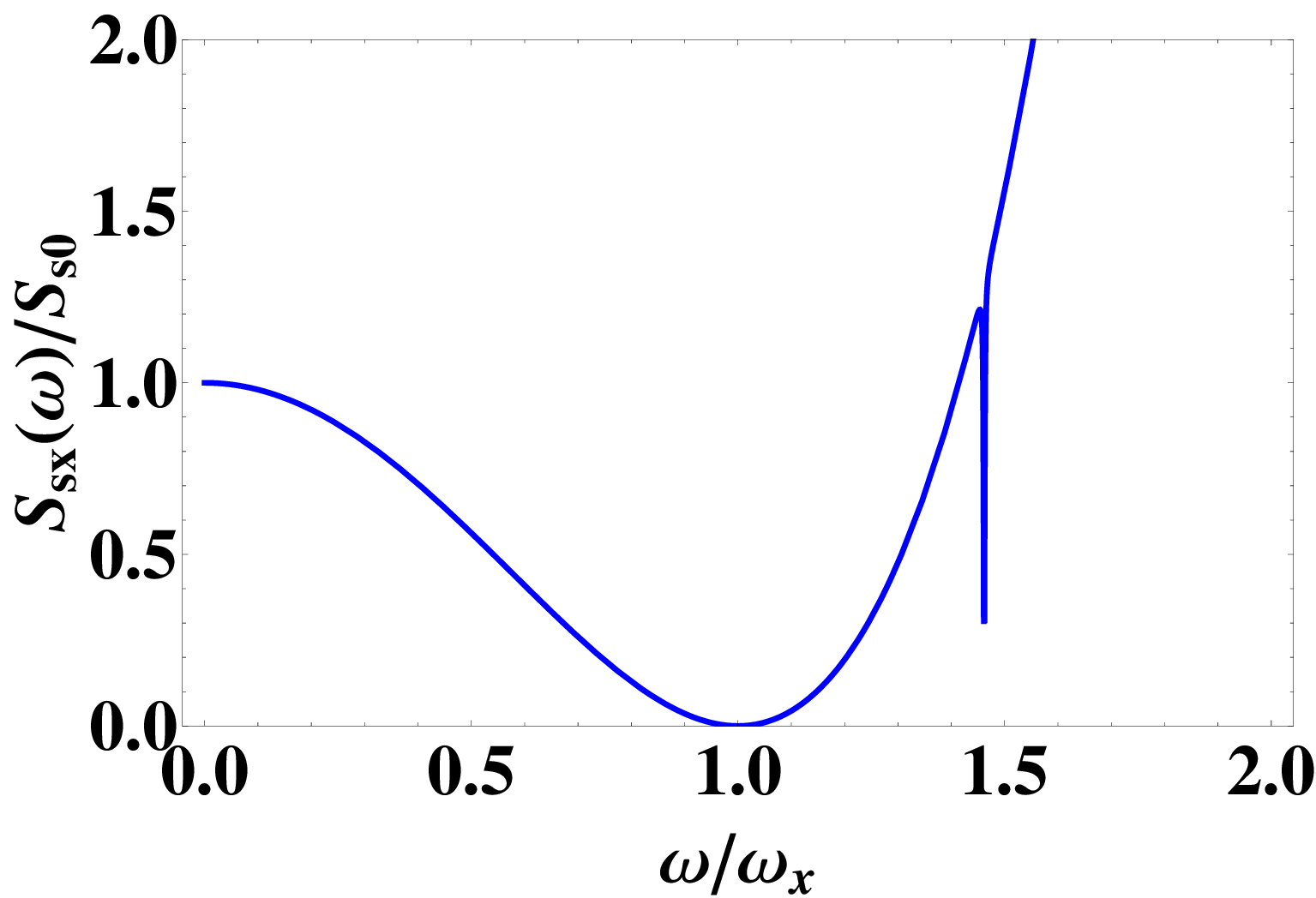}}%
\hspace{8pt}%
\subfigure[][]{%
\label{fig:fig9b}%
\includegraphics[height=3.8cm, width=3.9cm]{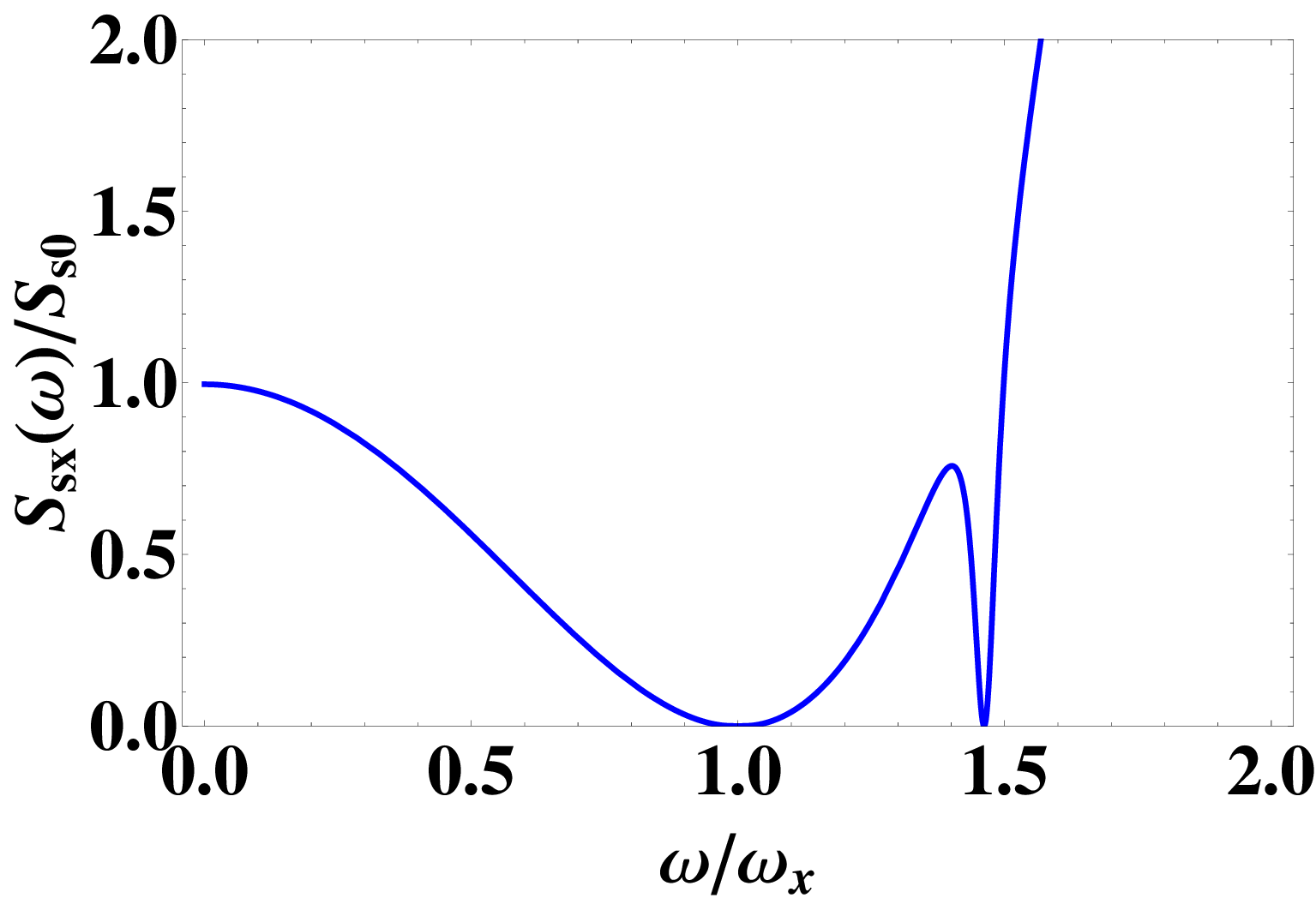}} \\
\hspace{8pt}%
\subfigure[][]{%
\label{fig:fig9c}%
\includegraphics[height=3.8cm, width=3.9cm]{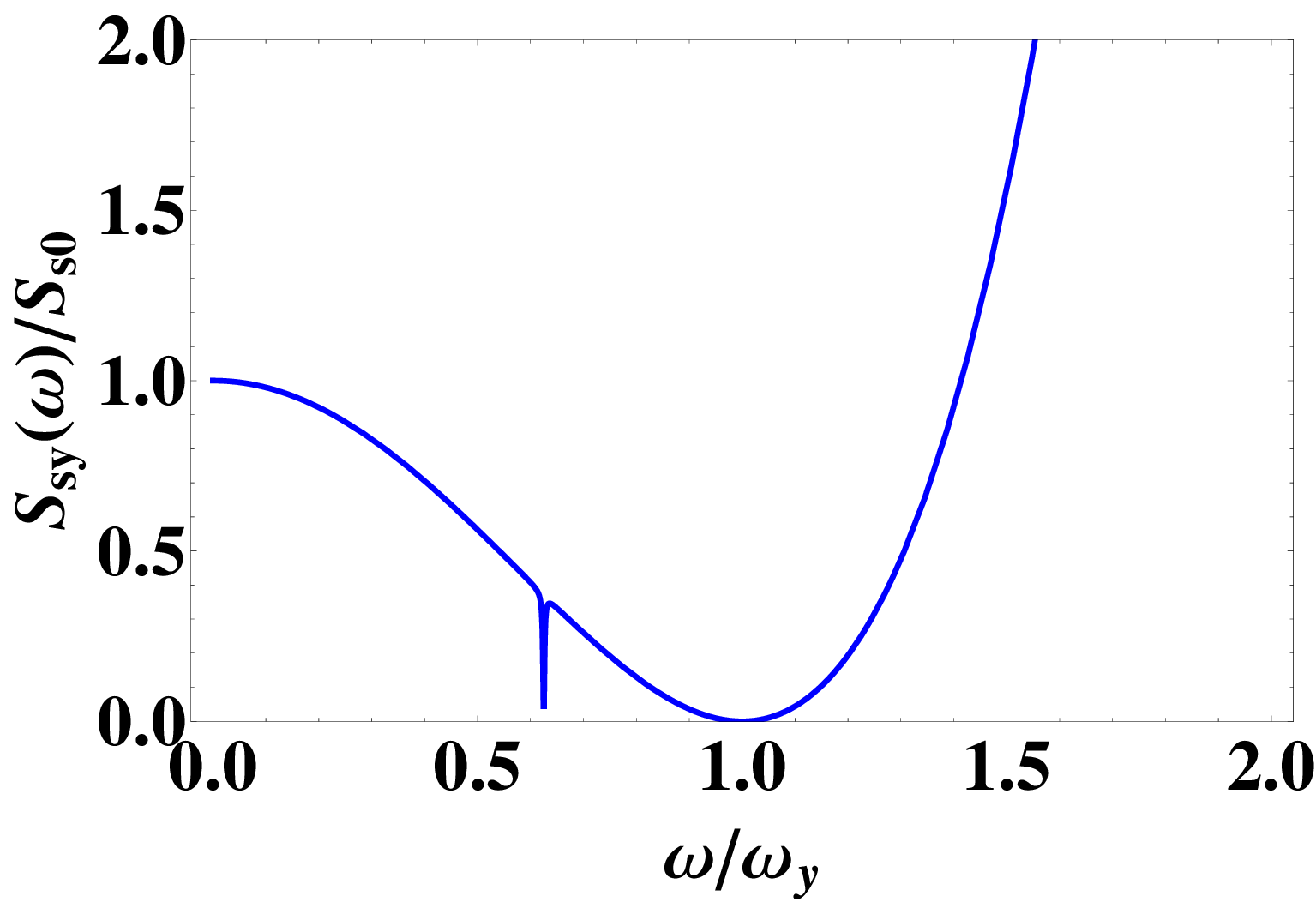}} 
\hspace{8pt}%
\subfigure[][]{%
\label{fig:fig9d}%
\includegraphics[height=3.8cm, width=3.9cm]{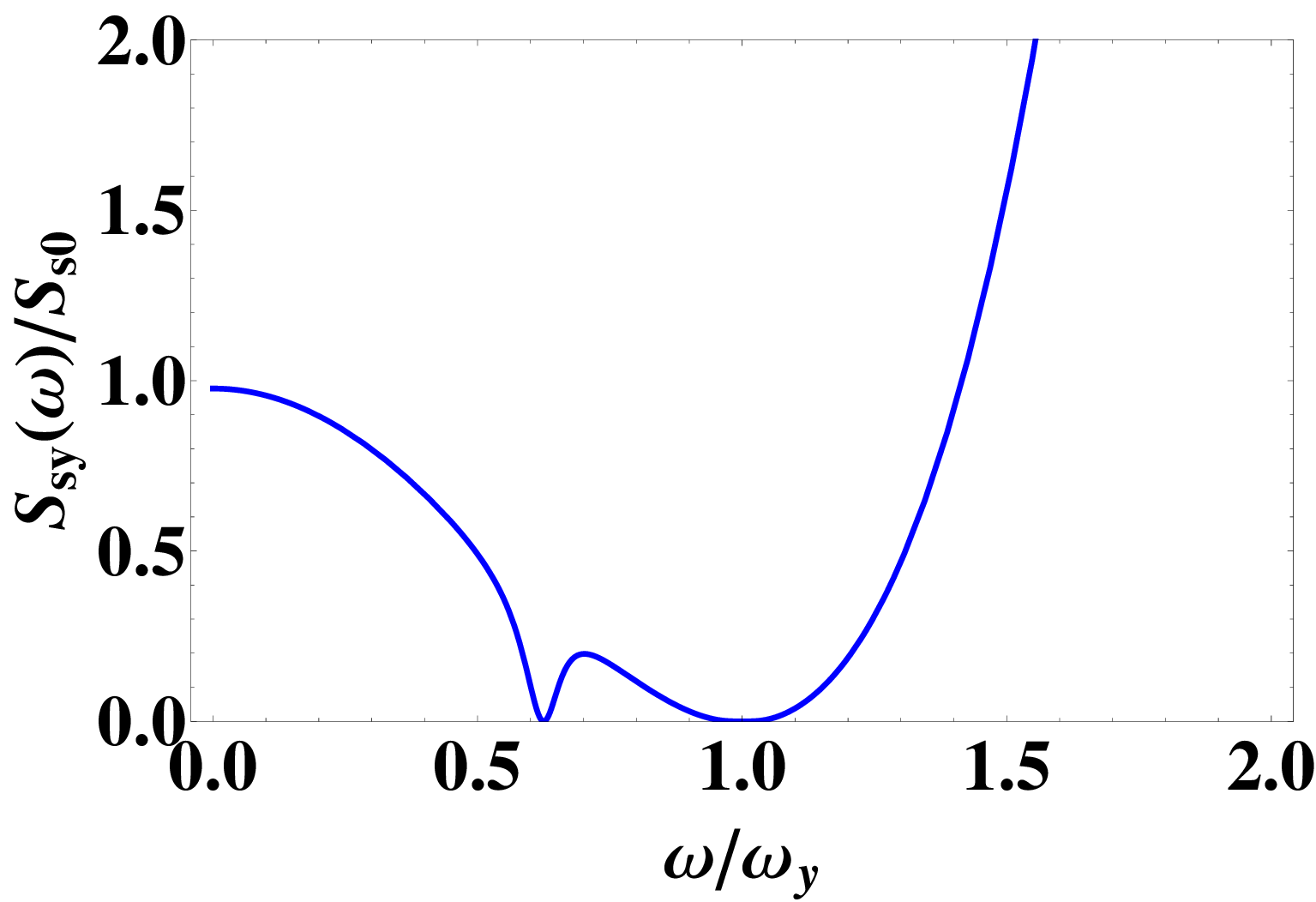}} 
 \caption[]{Shot noise force PSD as a function of frequency. Panels (a) and (b) show $S_{sx}(\omega)/S_{s0}$ of the $x$-mode for $\delta=10^{-5}$ and  $\delta=10^{-3}$, respectively. Panels (c) and (d) show $S_{sy}(\omega)/S_{s0}$ of the $y$-mode for $\delta=10^{-5}$ and  $\delta=10^{-3}$, respectively. The parameters are $\omega_{x}=130$ KHz, $\omega_{y}=160$ KHz, $\gamma_{gx}=0.06$ Hz, $\gamma_{cx}=10^{-4}$ Hz, $\gamma_{cy}=10^{-4}$ Hz, $\gamma_{gy}=0.06$ Hz, $\Gamma_{cx}=10^{-5}$ Hz, $\Gamma_{cy}=10^{-5}$ Hz, $\gamma_{ay}=0$ Hz, $D_{tx}=10$ KHz, $D_{ty}=10$ KHz, $\eta= 2 \times 10^{-9}$, $\phi = 5 \times 10^{16} \mathrm{photons/s}, $and T = 1K. The nanoparticle diameter $D = 100 $ nm and density $\rho=2200$ kg/m$^{3}$.}
 \label{fig:fig9}
 \end{figure}
%%%%%%%%%%%%%%%%%%%%%%%%%%%%%%%%%%%%%%%%%%%%%%%%%%%%%%%%%%%%%%%%%%%%%%
%%%%%%%%%%%%%%%%%%%%%%%%%%%%%%%%%%%%%%%%%%%%%%%%%%%%%%%%%%%%%%%%%%%%%%%%
%%%%%%%%%%%%%%%%%%%%%%%%%%%%%%%%%%%%%%%%%%%%%%%%%%%%%%%%%%%%%%%%%%%
\begin{align}
\label{Eq_13}
\chi_{j}(\omega)&=\frac{1+A_{v}(\omega)+B_{v}(\omega)}{m((\omega^{2}_{j}-\omega^{2})+ i\omega\Gamma_{j})}\,\,\,  \mathrm{for} ~ j \neq v \\
A_{v}(\omega)&=\frac{\kappa \delta}{2m((\omega^{2}_{v}-(\omega-\omega_{r})^{2})+ i(\omega-\omega_{r})\Gamma_{v})}, \nonumber \\
B_{v}(\omega)&=\frac{\kappa \delta}{2m((\omega^{2}_{v}-(\omega+\omega_{r})^{2})+ i(\omega+\omega_{r})\Gamma_{v})}. \nonumber
%\label{Eq_14}
\end{align}
%%%%%%%%%%%%%%%%%%%%%%%%%%%%%%%%%%%%%%%%%%%%%%%%%%%%%%%%%%%%%%%%%%%
where both $j ~\&~ v \in \{x,y\}$. We now include the effect of shot noise of the measured signal and write the positional power spectral density (PSD) for both the modes as 
%%%%%%%%%%%%%%%%%%%%%%%%%%%%%%%%%%%%%%%%%%%%%%%%%%%%%
\begin{align}
\label{Eq_15}
\langle|q_{j}(\omega)|^{2}\rangle=|\chi_{j}(\omega)|^{2}(S_{Tj}+S_{Hj}+S_{Cj})+\frac{l^{2}_{j}}{\eta^{2}\phi}.
\end{align}
%%%%%%%%%%%%%%%%%%%%%%%%%%%%%%%%%%%%%%%%%%%%%%%%%%%%%
where $\eta$, $\phi$ are the optomechanical coupling coefficient and photon flux \cite{Rodenburg}, respectively, and $l_{j}= \sqrt{\hbar/2m\omega_{j}}$ is the oscillator length, along with $ j \in \{x,y\}$. Further, the stochastic forces due to thermal, feedback heating and feedback cooling are given as $S_{Ti}=2m\gamma_{gj}K_{B}T$, $S_{Hj}=\hbar m\omega_{j}D_{tj}$, and $S_{Cj}=36\hbar m \omega_{j}\frac{\Gamma^{2}_{cj}}{\gamma_{cj}}[4\langle{N}_{j}\rangle^{2}+4\langle{N}_{j}\rangle+1]$, respectively.

Now from PSD equation as in Eq.~(\ref{Eq_15}), we can write the force PSD for each mode as
%%%%%%%%%%%%%%%%%%%%%%%%%%%%%%%%%%%%%%%%%%%%%%%%%%%%%\textcolor{blue}{Modified and to be Included:}\textcolor{red}{It is actually not a proper comparison but a comparison among themselves. What i means to say that if we consider the parameters in Fig.9 which is $\deta=10^{-5}$}
\begin{align}
\label{Eq_16}
\langle|F_{j}(\omega)|^{2}\rangle&=S_{Tj}+S_{Hj}+S_{Cj}+S_{sj}(\omega).
\end{align}
%%%%%%%%%%%%%%%%%%%%%%%%%%%%%%%%%%%%%%%%%%%%%%%%%%%%%\textcolor{red}{
where $S_{sj}(\omega)=S_{s0}/|\chi_{j}(\omega/\omega_{j})|^{2}$, with $S_{s0}=m^{2}l^{2}_{j}\omega^{4}_{j}/\eta^{2}\phi$ and $ j \in \{x,y\}$ \cite{Rodenburg}.

To obtain a minimum force sensitivity, we consider the case where the frequency dependent term $S_{sj}(\omega)$ is minimum. Hence, in this regard we numerically study $S_{sj}(\omega)$ and present the result in Fig.~\ref{fig:fig9}. It can be seen from Fig.~\ref{fig:fig9a} and Fig.~\ref{fig:fig9c}, which apply for low values of the coupling, that the force PSD ($S_{sj}(\omega)$) attains a minimum value around the oscillation frequencies for each mode. Considering optimal parameters as in Fig.~\ref{fig:fig9}, we find a minimum force sensitivity for the coupled system to be $\sqrt{\langle|F_{j}(\omega)|^{2}\rangle_{min}}$~$\approx$~$10^{-21}$$N/\sqrt{Hz}$. It is also interesting to notice that in the case of higher values of coupling, force noise PSD for each mode have minima at two points $\omega_{x}$ and $\omega_{y}+\omega_{r}$ for $x$-mode and $\omega_{y}$ and $\omega_{x}-\omega_{r}$ for $y$-mode, as shown in Fig.~\ref{fig:fig9b} and Fig.~\ref{fig:fig9d}, respectively. We find the minimum force sensitivity at other frequencies relevant to the minima of the force noise PSD also to be $\sqrt{\langle|F_{j}(\omega)|^{2}\rangle_{min}}$~$\approx$~$10^{-21}$$N/\sqrt{Hz}$ in the strong coupling regime.  

These features in the force PSD can be attributed to the phenomenon of mode-splitting which arises generally in a coupled system and has recently been specifically experimentally identified in a levitated nanoparticle \cite{Frimmer2}. Interestingly, these results imply that one can measure weak forces at quite different frequencies with high sensitivity by tuning the coupling in the system.  Hence, the coupled levitated naoparticle can be useful in designing tunable ultrasensitive sensors. 
%%%%%%%%%%%%%%%%%%%%%%%%%%%%%%%%%%%%%%%%%%%%%%%%%%%%%
\section{Conclusion}
\label{Con}
In conclusion, we have theoretically demonstrated PT symmetry, induced lasing and tunable ultrasensitive force sensing in a coupled-mode levitated system. We have used quantum Langevin equations along with phonon evolution and second order coherence to quantify the dynamics of the coupled system. We found that, in the regime where oscillation frequencies of both the mechanical modes as well as their strength of their coupling are very much larger than their respective damping rates, the position dynamics for both the modes show sustained modulation of oscillation, indicating PT symmetry. 

Although the system shows oscillation with constant amplitude, the dynamics of phonon population and tests of the second order coherence reveal both the modes to be in thermal rather than coherent states. We also show that when one of the modes is in a coherent state i.e. in lasing mode, the other mode also attains a coherent state due to the coupling, indicating induced lasing transfer. The lasing characteristics of the modes are validated by studying the phonon dynamics and the second order coherence for the coupled system. We also studied tunable force sensing in the coupled-mode levitated system and found sensitivity of the order of $zN/\sqrt{Hz}$. Our theoretical work suggests new possibilities for an optically levitated coupled-mode nanoparticle for coherent manipulation and force sensing.
%%%%%%%%%%%%%%%%%%%%%%%%%%%%%%%%%%%%%%%%%%%%%%%%%%%%
%%%%%%%%%%%%%%%%%%%%%%%%%%%%%%%%%%%%%%%%%%%%%%%%%%%%
\section{Acknowledgment}
This research is supported by the Office of Naval Research under Award Nos.
N00014-17-1-2291 and N00014-18-1-2370.
%%%%%%%%%%%%%%%%%%%%%%%%%%%%%%%%%%%%%%%%%%%%%%%%%%%%

%%

\end{document}